\documentclass [superscriptaddress,preprint,amsmath,amssymb,showkeys]{revtex4-1}
\usepackage[hidelinks]{hyperref}
\setcitestyle{super}
\usepackage{graphicx}
\usepackage{dcolumn}
\usepackage{bm}
\usepackage{color}
\usepackage{float}
\usepackage[utf8]{inputenc}
\usepackage[mathlines]{lineno}
\usepackage{multirow}
\usepackage{amsmath}
\usepackage{xr-hyper}
\usepackage{hyperref}
\usepackage{soul,xcolor}
\usepackage{natbib}
\usepackage{amsmath} 
\usepackage{xcolor}
\usepackage{xr}
\usepackage{chemformula}
\usepackage{siunitx}
\usepackage{physics}
\soulregister\cite7
\soulregister\ref7
\setstcolor{blue}
\newcommand{\doHMN}[2]{%
  \begingroup\lccode`~=`#1
  \lowercase{\endgroup\let~}#2%
  \mathcode`#1="8000
}

\makeatletter
\newcommand*{\addFileDependency}[1]{
  \typeout{(#1)}
  \@addtofilelist{#1}
  \IfFileExists{#1}{}{\typeout{No file #1.}}
}
\makeatother

\newcommand{\angstrom}{\text{\normalfont\AA}}

\begin{document}
\title{$\emph{Ab initio}$ study of proton-exchanged LiNbO$_3$(I): Structural, thermodynamic, dielectric, and optical properties}

\author{Lingyuan Gao}
\thanks{These authors contributed equally}
\affiliation{Department of Chemistry, University of Pennsylvania, Philadelphia, Pennsylvania 19104--6323, USA}
\author{Robert B. Wexler}
\thanks{These authors contributed equally}
\affiliation{Department of Chemistry, University of Pennsylvania, Philadelphia, Pennsylvania 19104--6323, USA}
\author{Ruixiang Fei}
\thanks{These authors contributed equally}
\affiliation{Department of Chemistry, University of Pennsylvania, Philadelphia, Pennsylvania 19104--6323, USA}
\author{Andrew M. Rappe}
\email{rappe@sas.upenn.edu}
\affiliation{Department of Chemistry, University of Pennsylvania, Philadelphia, Pennsylvania 19104--6323, USA}

\date{\today}

\begin{abstract}
Using first principles calculations, we study the ground-state structure of bulk proton-exchanged lithium niobate, which is also called hydrogen niobate and is widely used in waveguides. Thermodynamics helps to establish the most favorable nonpolar surface as well as the water-deficient and water-rich phases under different ambient conditions, which we refer to as ``dehydrated'' and ``rehydrated'' phases, respectively. We compute the low-frequency dielectric response and the optical refractive indices of hydrogen niobate in different phases. The dielectric constant is greatly enhanced compared to lithium niobate. At shorter wavelengths, the refractive indices vary between each phase and have a sharp contrast to lithium niobate. Our study characterizes the structures and thermal instabilities of this compound and reveals its excellent dielectric and optical properties, which can be important in the future application in waveguides.  
\end{abstract}


\maketitle
\def\thefootnote{+}\footnotetext{These authors contributed equally to this work}\def\thefootnote{\arabic{footnote}}

\section*{Introduction} 
Lithium niobate LiNbO$_3$ (LNO) is a human-made ferroelectric material~\cite{Carruthers71p1846}. With a high phase transition temperature close to its melting point (1480 K), it exhibits excellent piezoelectric, electro-optic, photorefractive, and nonlinear optical properties. 
These advantages make LNO widely used in optical phase modulators~\cite{nishihara00p26, Wooten00p69}, holographic memory storage~\cite{Chen95p33, Miguel-Sanz65p165101}, and nonlinear frequency converters~\cite{Zhu92p904,Pruneri95p2375}.
Owing to its large electro-optic coefficient, high refractive index, and wide frequency range with high optical transparency, nowadays LNO has become one of the most promising platforms for integrated photonics~\cite{Kosters09p510}. 

Currently, there are two main methods to integrate LNO-based optical waveguides: (1) Titanium in-diffusion~\cite{Schmidt74p458,arizmendi2004photonic}, and (2) proton exchange.
The first approach has advantages of small effective diffusion length and good confinement of both ordinary and extraordinary waves (increase of refractive indices $\delta n \approx 0.04$)~\cite{Schmidt74p458, Neyer79p256, Fukuma80p591}.
Alternatively, proton exchange is more straightforward in that by immersing LNO in an acid such as benzoic acid, Li$^{+}$ ions diffuse out of the crystal while H$^{+}$ ions diffuse in~\cite{Jackel82p607}.
The replacement of Li$^{+}$ with H$^{+}$ usually takes place up to several microns from the surface of LNO~\cite{Cabrera96p349}, and the change of the composition in this surface layer induces a jump of refractive index relative to the part below.
Compared to titanium in-diffusion, proton-exchange gives a higher refractive index change for the extraordinary wave ($\delta n_{e} \approx 0.12$)~\cite{Jackel82p607,Rice86p188}.
The proton exchange is usually followed by an annealing step at a controllable duration and temperature, called as ``annealing proton exchange''(APE), which can restore electro-optical effect and nonlinear coefficients and lower propagation loss~\cite{korkishko96p7056,Korkishko96p175}.
Sometimes, a reverse proton exchange (RPE) procedure is also adopted by re-exchanging the protons near the surface back to Li$^{+}$ ions, therby burying exchanged waveguides in eutectic or pure LNO~\cite{korkishko95p149,Korkishko96p187}. RPE can increase depth index profile symmetrization and reduce fiber waveguide coupling loss~\cite{Korkishko96p175,korkishko98p1838}.  

Generally, proton exchange can be viewed as a chemical reaction~\cite{Ohsaka01p2141}, expressed as:
\begin{equation}
 x \text{H}^{+} + \text{LiNbO}_3 \rightleftharpoons \text{H}_x  \text{Li}_{1-x} \text{NbO}_3+ x \text{Li}^{+}.
\end{equation}
H$_{x}$Li$_{1-x}$NbO$_3$ is a complex product, and there can be up to seven different crystallographic phases, highly dependent on the orientation of crystal cut (X-, Y-, or Z-) and the proton concentration $x$~\cite{Fedorov94p243, fedorov95p216, Korkishko96p175, Korkishko96p187,korkishko96p7056,Korkishko97p188}.
As a function of $x$ in the exchange layer, lattice parameters can change abruptly and there can be sudden jumps between different crystal phases~\cite{Fedorov94p243, fedorov95p216, Korkishko96p175, Korkishko96p187,korkishko96p7056,Korkishko97p188}.
This induces an abrupt change of local refractive index and is detrimental to the performance of the waveguide. Therefore, it is vital to characterize the structural and optical properties of H$_{x}$Li$_{1-x}$NbO$_3$.

In this work, instead of examining all the possible phases with different proton concentration $x$, we focus on hydrogen niobate $\rm HNbO_{3}$ (HNO), which can be synthesized from LiNbO$_3$ via ion exchange when Li$^{+}$ ions are fully replaced by H$^{+}$ ions~\cite{Rice82p308}.
The forward conversion from LNO to HNO is more complete than the backward conversion from HNO to LNO in concentrated acid treatment~\cite{Ohsaka01p2141}. X-ray diffraction shows that $\rm HNbO_3$ is a cubic perovskite with corner-sharing $\rm NbO_6$ units and the refined lattice constant is $a$ = 7.645 $\angstrom$~\cite{Fourquet83p1011,Pokrovskii00P890,Kalabin03p140}.
Nevertheless, the positions of H atoms are not well located, and previous studies showed that they are statistically distributed on two types of positions~\cite{Fourquet83p1011}. Nuclear magnetic resonance (NMR) spectra also indicated that protons can easily jump from one oxygen to another unbonded oxygen~\cite{Weller85p139}.

Using first principles calculations, we first search for the ground state structure using an automatic algorithm. As we find, the ground-state structure is highly polarized, with all OH$^{-}$ dipoles aligned along the same direction. These OH$^{-}$ dipoles can be flipped with an external electric field at a small energy cost.
From the predicted bulk structure, we establish the most stable nonpolar HNbO$_3$ surface and identify the corresponding passivation layers and adsorbates in ambient environment. The following thermodynamic studies show that water can be easily desorbed and reabsorbed into HNO under different ambient conditions, which leads to the formation of the water-deficient (``dehydrtaed'') or water-rich (``rehyrated'') phases. 
We finally study the dynamic low-frequency 
dielectric response and optical refractive 
indices of HNO in different phases. The dielectric constant is almost doubled in polar HNO compared to that of LNO, and the refractive indices also differ from LNO and vary between different phases.

\section{structure exploration of hydrogen niobate} 
\subsection{Ground-state structure of bulk hydrogen niobate}
The experimentally observed crystal structure of HNO
is shown in Fig. 1(a)~\cite{Fourquet83p1011}. The cubic unit cell contains 8 NbO$_6$ octahedra and the tilting pattern is $a^{0} a^{0}c^{+}$. We note that this is a conventional cell, and the primitive cell has 20 atoms.
48 available sites are available to be occupied by 8 H atoms, which generates 377,348,994 configurations in total. With such a huge number, we carry out a pseudorandom approach with the help of density functional theory (DFT) to search for the ground state zero-temperature structure.

\begin{figure}
\centering
\includegraphics[width=8 cm]{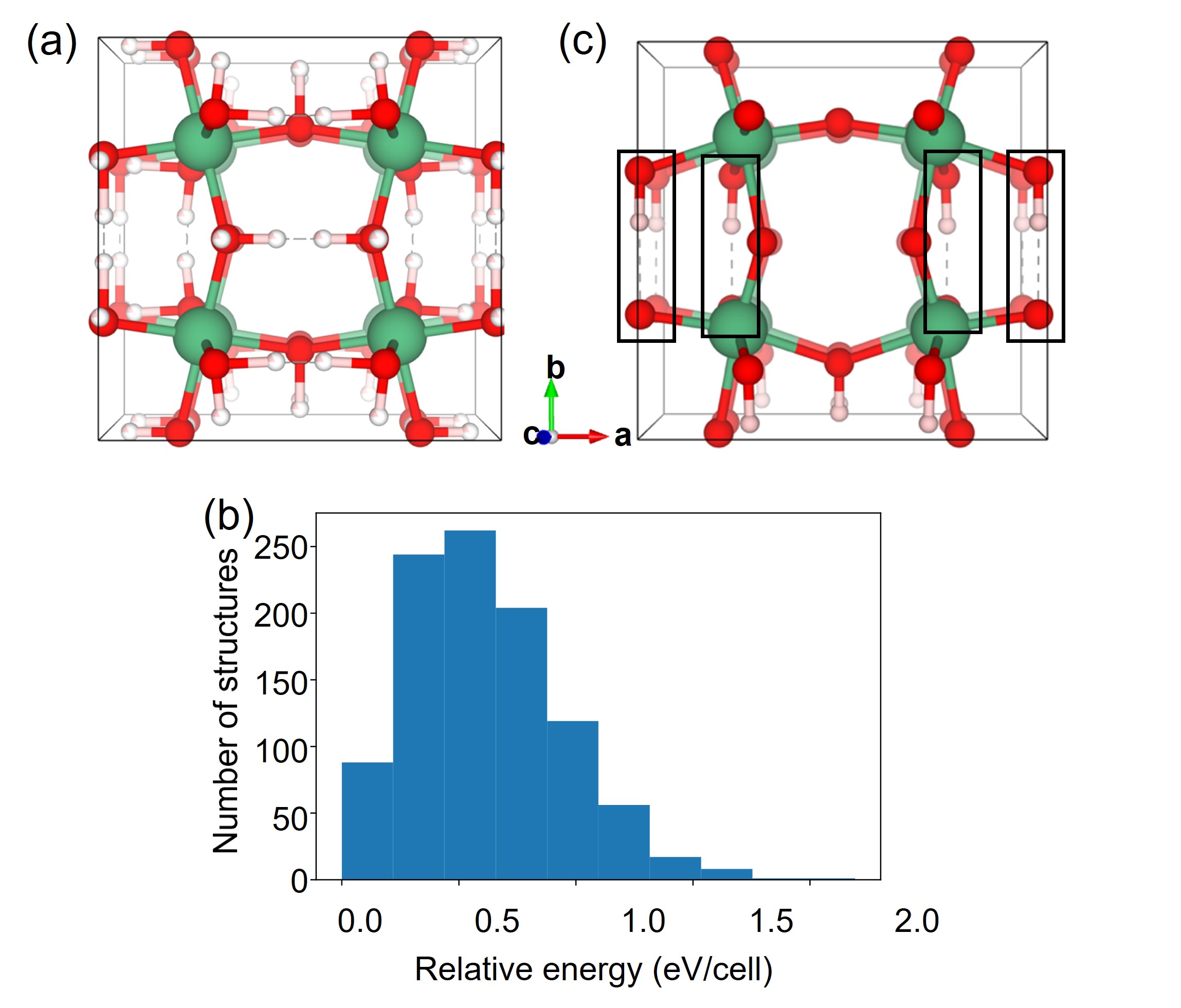}
\caption{\label{fig:bulk}
(a) Experimental crystal structure of HNO~\cite{Fourquet83p1011}. White, green, and red denote H, Nb, and O atoms, respectively. There are 48 available positions to occupy 8 H atoms. (b) Count distribution of relative energies of 1,000 candidate structures. (c) Computed ground-state structure of HNO. Compared with (a), we see that particular positions have been determined by energy minimization for the 8 H atoms. 4 O-O bridges are highlighted by black squares. 4 protons around $y$ = 0.5 are close to the mid-point of these O-O bridges.} 
\end{figure}

We establish three rules as the constraints of our searching algorithm: (1) The number of H bonded to each O atom should be no more than one. This corresponds to the condition that the formation of H$_2$O is unfavorable.
(2) H-H distances should be longer than 1$\angstrom$, so as to avoid the unfavorable formation of H$_2$ in HNO. (3) The number of OH$^{-}$ bonded to each Nb atom should not exceed three, to ensure that each Nb is locally charge balanced.
With these constraints, we randomly generate 1000 structures by dropping 8 H atoms on available positions in the cubic cell with the experimental lattice parameters (Fig.~\ref{fig:bulk}(a)).
We then conduct geometric relaxations for all 1000 structures including changing the shape and volume of the cell and computing their total energies.
Our DFT calculations are performed using the \textsc{Quantum Espresso} package~\cite{Giannozzi09p395502}, and we use norm-conserving pseudopotentials generated by the OPIUM package~\cite{Rappe90p1227}. The generalized gradient approximation (GGA) exchange-correlation functional~\cite{Perdew96p3865} is adopted.
   
\begin{figure}
\centering
\includegraphics[width=8 cm]{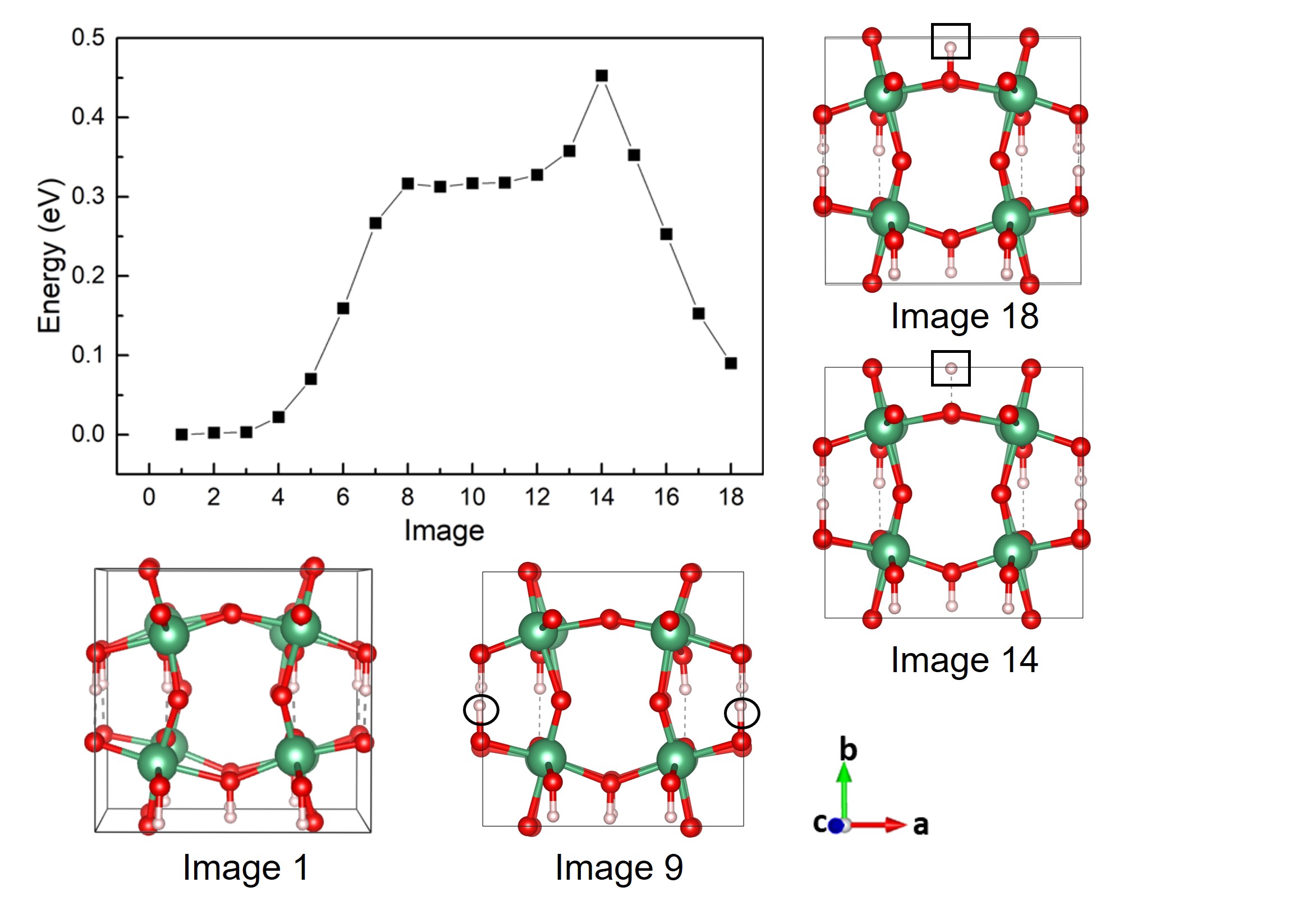}
\caption{\label{fig:OHflip}
The energy barrier for flipping OH$^{-}$ dipoles by the NEB calculation. Structures of one- and two-OH$^{-}$ flipped structures are given (Image 9 and image 18, respectively). Flipping OH$^{-}$ dipoles can be accomplished by displacing H atoms from one side to the other side of the O-O bridge along the $b$ ([010]) direction. In Image 9, black circles highlight the displaced proton close to $y$ = 0.5. In image 14 and 18, black square highlights the displaced proton close to $y$ = 1. } 
\end{figure}

Fig.~\ref{fig:bulk}(b) shows the count distribution of relative energies of all 1,000 structures. The total energy per conventional cell varies in a 1.8 eV/cell range with the median energy of 0.59 eV.
Energetically, many configurations are close to the ground-state structure. The lowest-energy structure of HNO is shown in Fig.~\ref{fig:bulk}(c). It is triclinic but close to tetragonal, with $a$ = 7.67 $\angstrom, b = 7.33 $\angstrom, $c$ = 7.68 $\angstrom$, $\alpha$ = 90.002$^{\circ}$,  $\beta$ = 89.722$^{\circ}$,  and $\gamma$ = 90.000$^{\circ}$.
The tilting pattern is $a^0a^0c^+$, consistent with the experimentally observed structure. In each NbO layer, 4 H atoms are attached on 4 different O atoms in a symmetric pattern, and each proton is close to the mid-point of the O-O bridge along $b$ ([010]) direction.
In this way, eight OH$^{-}$ dipoles are aligned in parallel and maximize the polarization of the system. However, since many configurations are thermally accessible according to Fig.~\ref{fig:bulk}(b), the OH$^{-}$ dipoles can be very disordered, and different orientations of OH$^{-}$ will induce a change of the polarization of the system.
By carrying out nudged elastic band (NEB) calculations, we calculate the energy barrier for flipping OH$^{-}$ dipoles, which can be completed by displacing a H atom from one side of the O-O bridge to the other side. As shown in Fig.~\ref{fig:OHflip}, to flip the first OH$^{-}$ dipole (at $y$ = 0.5) requires an energy about 0.3 eV (Image 9), and then to flip the second OH$^{-}$ dipole (at $y$ = 1) requires an energy about 0.45 eV (Image 18).
The energy barrier is much smaller compared to the energy of H-bond and OH-bond, and that indicates that the polarization of HNO can be easily switched with a small electric field.
Comparing the total energies, we find that one- and two-OH$^{-}$ flipped structures are 0.3 eV and 50 meV higher than the ground-state polar HNO. This demonstrates that HNOs with various OH$^{-}$ orientations and polarizations widely exist in nature.

\subsection{Stable nonpolar surface of hydrogen niobate}
\begin{figure*}
\centering
\includegraphics[width=16 cm]{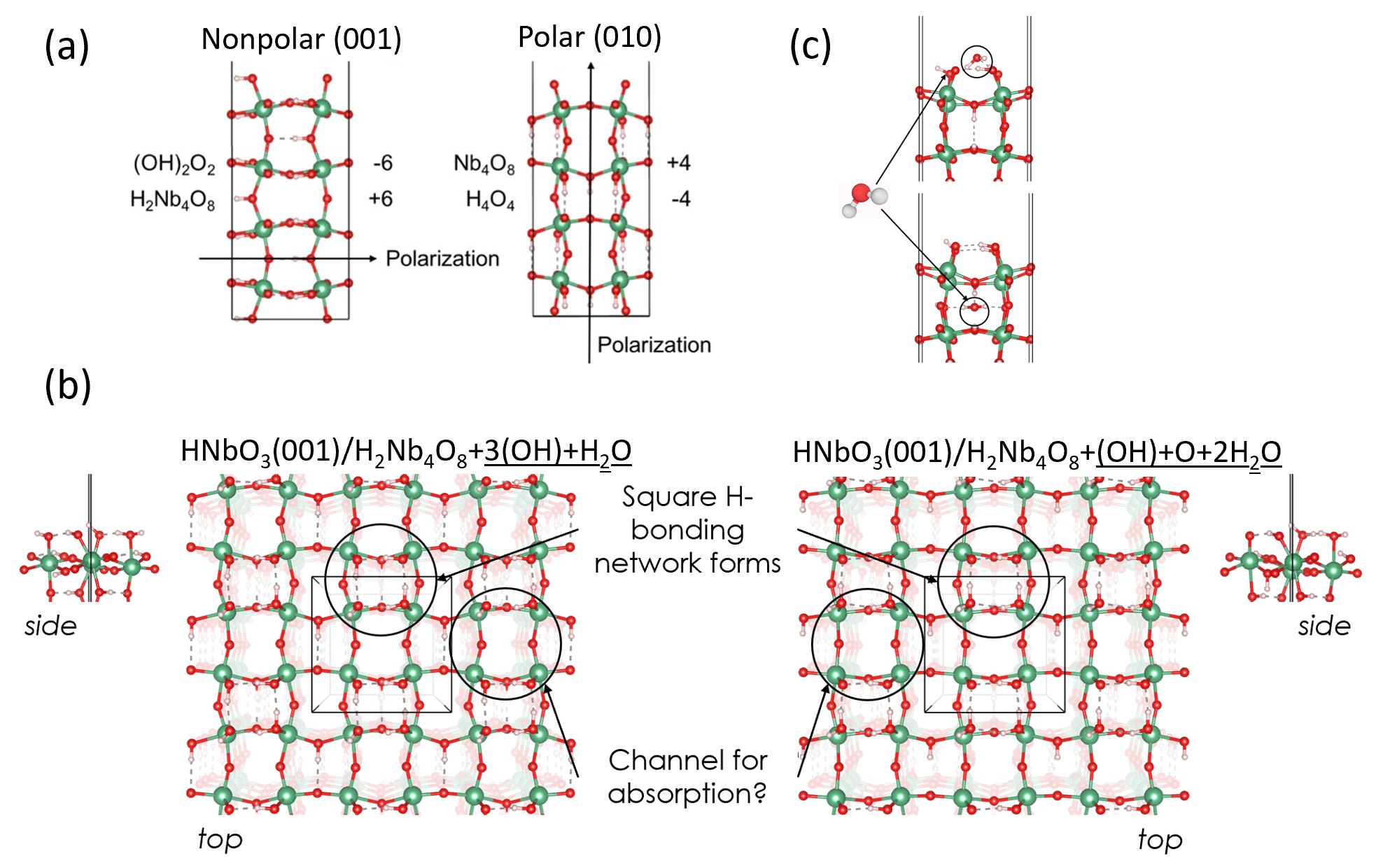}
\caption{\label{fig:surface}
(a) Slab models of nonpolar (001) and polar (010) HNO surfaces. Components and charges of each layer are different. (b) Two most stable HNO (001) passivated layers with different adsorbates. The planarities are also shown from the side view. (c) Illustration of an adsorption of an extra H$_2$O on the surface and absorption of an extra H$_2$O through HNO(001) surface. } 
\end{figure*}

To date, there have not been many studies about HNO surfaces. The Bravais-Friedel-Donnay-Harker (BFDH) theory states that the morphological importance of a particular facet is proportional to the interplanar spacing along the direction normal to this facet~\cite{Bravais86, Donnay37p446, Friedel07p326}.
Since in HNO three lattice constants are similar, the probabilities of the formation of ($\pm$100), (0$\pm$10), and (00$\pm$1) surfaces are approximately equal.
Also, since $a$ and $c$ directions are symmetrically equivalent, we only consider (0$\pm$10) and (00$\pm$1) surfaces.
Fig.~\ref{fig:surface}(a) shows that the (001) and (010) surfaces are different in terms of their polarities: The OH$^{-}$ dipoles in the (001) surface point to the left within the surface plane, while dipoles in the (010) surface point downward normal to the surface.
[(OH)$_2$O$_2$]$^{-6}$ and [H$_2$Nb$_4$O$_8$]$^{+6}$ layers are alternatingly stacked along the [001] direction, while [H$_4$O$_4$]$^{-4}$ and [Nb$_4$O$_8$]$^{+4}$ layers are stacked along the [010] direction. Below, we only consider the nonpolar (001) surface. 

As shown in Fig.~\ref{fig:surface}(a), the dangling OH$^{-}$ dipoles and O$^{2-}$ ions on the top of (001) surface can induce dipoles that destabilize the surface. These surface dipoles can be chemically passivated by ambient H$_2$O and O$_2$. After considering thermodynamics of 25 possible passivation schemes, we identify two most stable surfaces:   
\begin{flalign}
      & \rm{ HNbO_3(001): H_2 Nb_4 O_8+OH^{-}+O^{2-}+2H_2O },\nonumber \\
      &  \rm{ HNbO_3(001): H_2 Nb_4 O_8+3OH^{-}+H_2O }.
\end{flalign}
Here, H$_2$Nb$_4$O$_8$ refers to the pre-passivation surface layer, and adsorbates include OH$^{-}$, O$^{2-}$, and H$_2$O. Structures of the two passivated surface layers are illustrated in Fig.~\ref{fig:surface}(b).
Adsorbates sitting on Nb$^{5+}$ sites form square H-bonding networks, surrounded by Nb$_4$ squares.
Binding energies of H$_2$O to the surface are 1.08 eV and 1.24 eV for 2H$_2$O and 3H$_2$O adsorbed, respectively.
Large binding energies suggest H$_2$O binds strongly to the (001) surface, and the Nb$^{5+}$ sites without OH$^{-}$ and O$^{2-}$ should be saturated with H$_2$O.
As a result, these sites are possible channels for the absorption and desorption of water.
These two configurations have the lowest surface energies, which are 0.18 $\rm J/m^2$ and 0.17 $\rm J/m^2$ for 2H$_2$O and 3H$_2$O adsorbed, respectively. As a comparison, for niobate-deficient ($\rm HNbO_3(001): H_2 Nb_4O_8 - NbO$) layer, the surface energy is 0.3 $\rm J/m^2$.  

While these two surfaces are thermodynamically preferred, other surfaces can also be accessed kinetically, albeit transiently.
Here, we considered two (of many) possible classes of kinetic events: an extra H$_2$O adsorption to the adsorbates on top of $\rm H_2 Nb_4O_8$, and H$_2$O absorption in the bulk.
As shown in Fig.~\ref{fig:surface}(c), the H-bonding network from the adsorbed OH$^{-}$ and H$_2$O provides an adsorption site for an extra H$_2$O, and the adsorption energy is 0.66 eV.
If we leave the H$_2$O beneath the surface layer $\rm H_2 Nb_4O_8$, representing the H$_2$O is transported through the surface and fully absorbed in the bulk region, the absorption energy is -1.28 eV. The negative number suggests that the surface is not water permeable, and H$_2$O molecules prefer to be adsorbed on the surface.

\section{Thermodynamic stability of hydrogen niobate} 
\begin{figure}[H]
\centering
\includegraphics[width=8 cm]{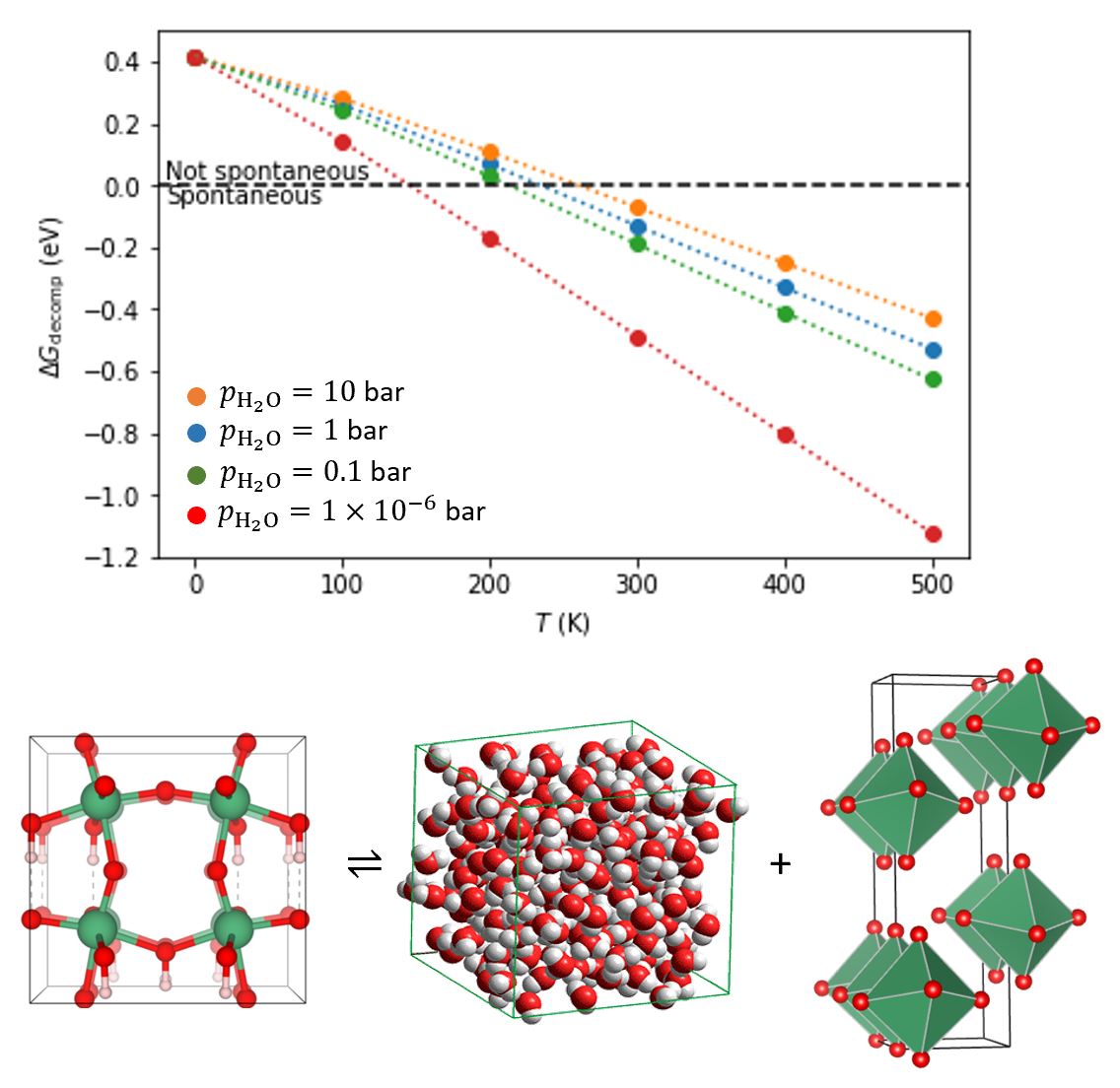}
\caption{\label{fig:decompose}
Gibbs free energy diagram of HNO for the complete dehydration into Nb$_2$O$_5$ under different temperatures and pressures of H$_2$O. The process is illustrated on the bottom.} 
\end{figure}

In HNO, apart from adsorption, H$_2$O can also escape from HNO via the pathway
\begin{equation}
    \rm 2HNbO_3 \rightleftharpoons  H_2 O+Nb_2 O_5.
\end{equation}
Nb$_2$O$_5$, a stable bulk niobium oxide, is the other product of this decomposition. By releasing H$_2$O, this can be thought as a dehydration process.
Figure~\ref{fig:decompose} shows the Gibbs free energy $G = H -TS$ of the complete dehydration as a function of temperature and pressure of H$_2$O.
For all pressures, when the temperature increases, dehydration becomes more spontaneous. The transition between non-spontaneous and spontaneous dehydration occurs between 150 K and 250 K, depending weakly on water vapor.
The figure reveals that near or above room temperature, HNO is expected to decompose spontaneously. In the transition region, the contribution from entropy $TS$ is more significant than the contribution from enthalpy $H$.
When the pressure of H$_2$O decreases from 1 bar to 1$\times$10$^{-6}$ bar, the transition temperature will also decrease, in line with Le Ch$\rm{\hat{a}}$telier's principle.

\begin{figure*}
\centering
\includegraphics[width=12 cm]{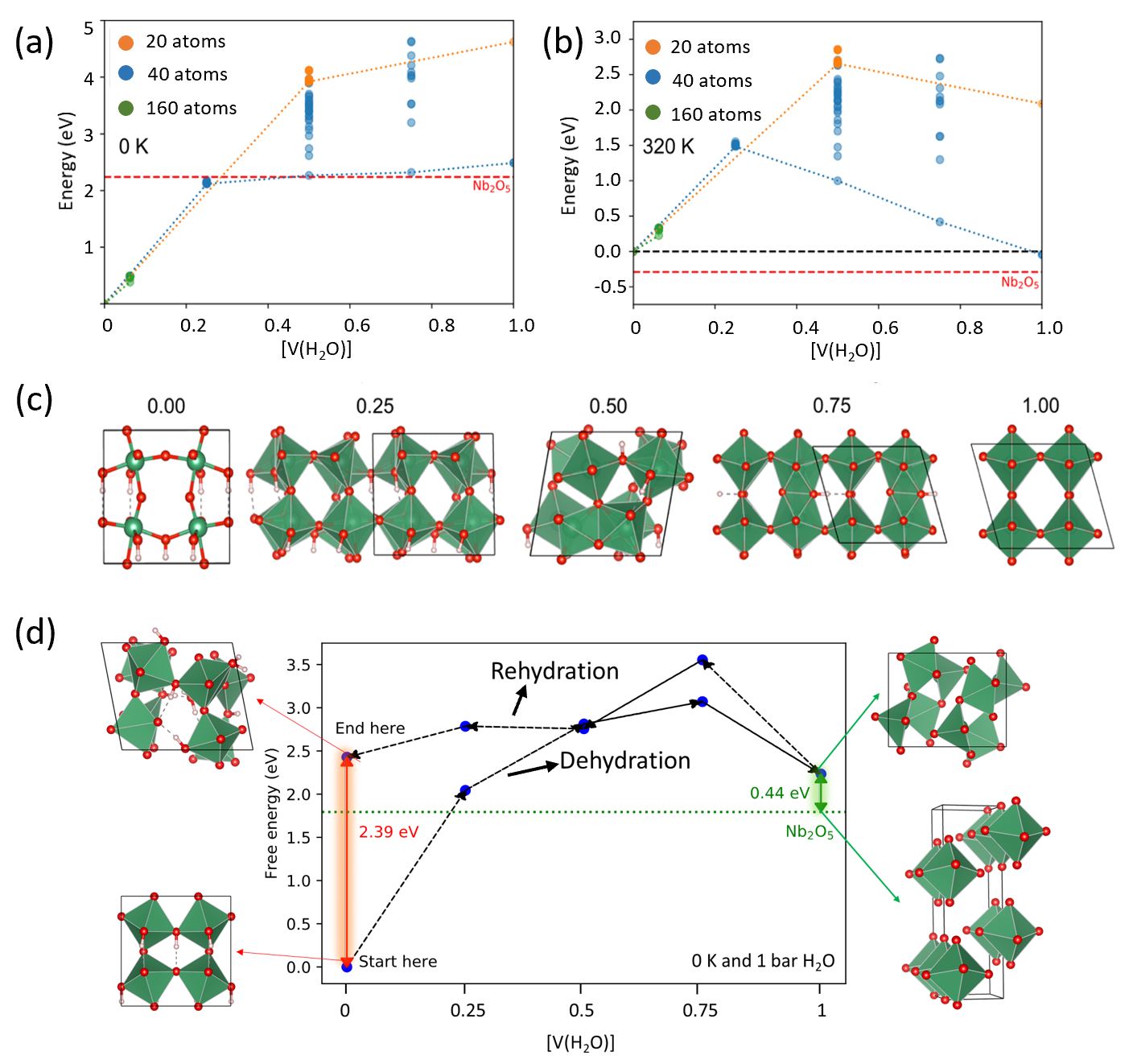}
\caption{\label{fig:dehydrate}
(a) Energy profile for partial HNO dehydration at 1 bar and 0 K. Orange, blue, and green correspond to a 20-atom primitive cell, 40-atom conventional cell, and 160-atom 2$\times$2$\times$2 supercell, respectively. Dotted lines connect minimum energy points at different $x$. As a reference, the energy of bulk Nb$_2$O$_5$ is denoted with the red dashed line.
(b) Similar to (a), but the energy profile is at room temperature 320 K. 
(c) Structural changes of HNO at different partial dehydration conditions at 0 K.
(d) A complete dehydration-rehydration hysteresis of HNO in a 40-atom conventional cell when the water is added back to the final dehydrated structure  at 320 K. Rehydration gives a different HNO structure at the end from the ground state. } 
\end{figure*}

At equilibrium, the dehydration process can be partial, i.e.,
\begin{equation}
   \text{HNbO}_3\rightleftharpoons 2x\text{H}_2\text{O}+\text{H}_{1-x} \text{NbO}_{3-2x},
\end{equation}
where $0 < x < 1.0$. Due to the limit of the size of the supercell, we model a few possible values of $x$.
For example, we can model $x\in$ \{0, 0.50, 1.0\} in a 20-atom primitive cell, and we can model $x\in$ \{0, 0.25, 0.50, 0.75, 1.0\} in a 40-atom conventional cell. In a 160-atom 2$\times$2$\times$2 supercell, we can model $0.0 < x <1.0$ in increments of 0.625.
The results of partial dehydration calculations for different $x$ and temperatures are shown in Figure~\ref{fig:dehydrate}.
At 0 K (Fig.~\ref{fig:dehydrate}(a)), the partial dehydration of HNO is unfavorable, as more than 2 eV per cell is needed to achieve 25\% dehydration.
As illustrated in Fig.~\ref{fig:dehydrate}(c), the structure remains approximately the same when $x = 0.25$ and the tilting pattern is mostly retained.
When $x > 25$\%, dehydration is more favorable with a lower cost of energy, and structural change is more dramatic with an orthorhombic-monoclinic phase transition at $x = 0.50$.
Edge-sharing Nb-O polyhedra are formed, resembling bulk Nb$_2$O$_5$. The final structure at $x = 1.0$ is slightly higher in energy than bulk Nb$_2$O$_5$, and that is likely due to the absence of thermal energy for overcoming kinetic barriers to Nb$_2$O$_5$.
At 320 K (Fig.~\ref{fig:dehydrate}(b)), the partial dehydration is still unfavorable from the intact HNO to 25\% dehydration but now it requires 0.5 eV less compared to the dehydration at 0 K. Beyond 25\%, dehydration is downhill in energy, indicating that once the system reaches this turning point, it will be spontaneously decomposed to Nb$_2$O$_5$. 

Next, we add the water back to the final dehydrated loop to form a complete dehydrate-rehydrate hysteresis.
Results are shown in Fig.~\ref{fig:dehydrate}(d); clearly, the rehydration pathway gives different products from the dehydration pathway. The HNO structure can be gradually recovered, and the final rehydrated structure is 2.5 eV higher compared to the polar HNO, demonstrating that the rehydrated HNO is less stable. 

\begin{figure*}
\centering
\includegraphics[width=8 cm]{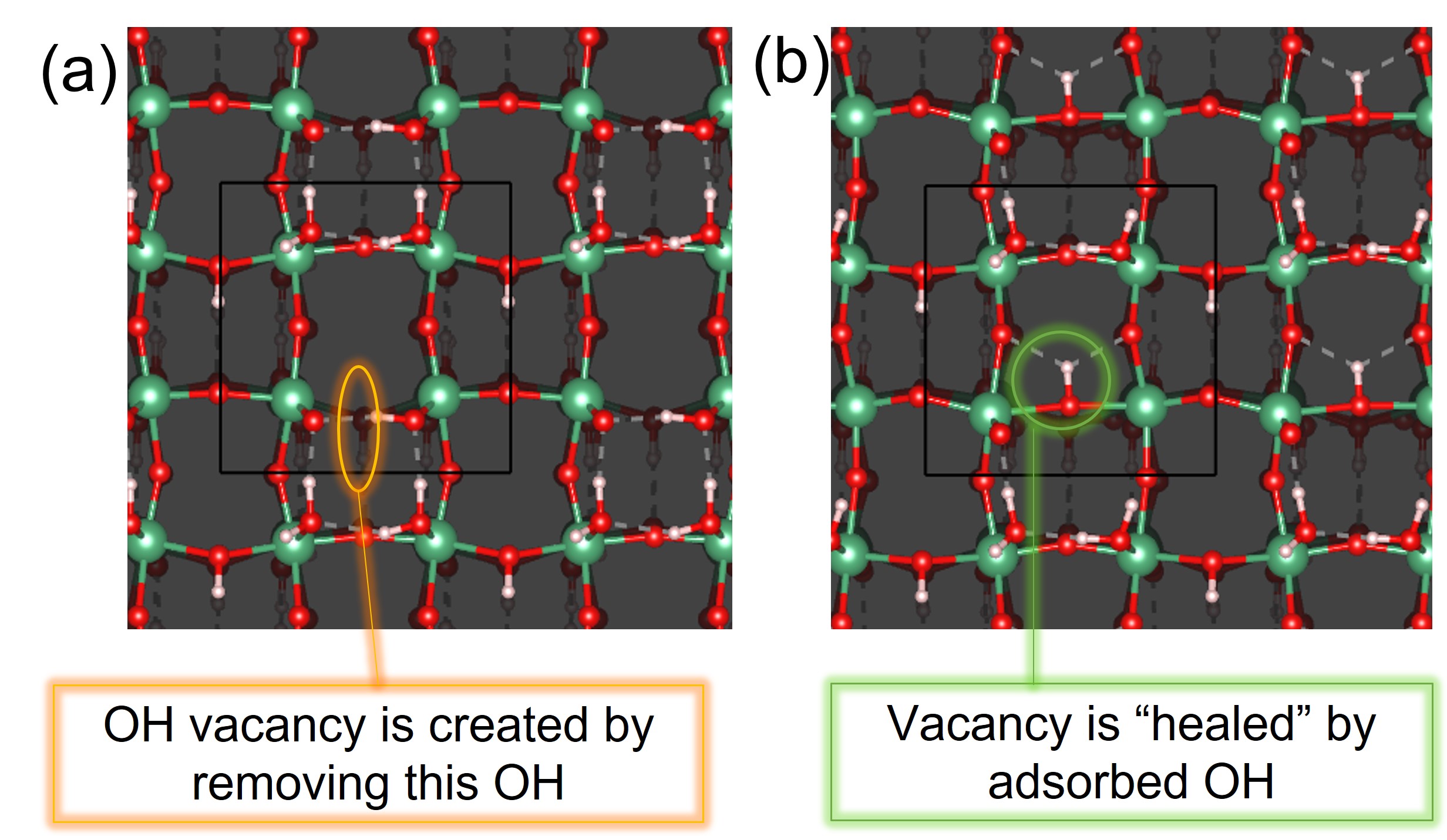}
\caption{\label{fig:surfacedehydrate}
(a)The most stable dehydrated nonpolar HNO (001) surface. An H$_2$O vacancy is created by removing an OH$^{-}$ highlighted by the orange circle, and a H atom below the black plane. (b)After performing structural relaxation, OH vacancy is healed by the migration of adsorbed OH$^{-}$ from the nearby Nb$^{5+}$ site, and H vacancy is also healed by extra H ions. } 
\end{figure*}

To make the analysis more complete, we also consider the dehydration at the surface of HNO. Fig.~\ref{fig:surfacedehydrate}(a) shows the top view of the most stable dehydrated nonpolar HNO(001) surface, built from the 2H$_2$O-adsorbed surface shown in Fig.~\ref{fig:surface}(b).
H$_2$O vacancies are created by removing OH$^{-}$ ions on the top H$_2$Nb$_4$O$_8$ layer (indicated by the orange circle) and removing hydrogen atoms below (beneath the black plane, not visible here).
In fact, as we will show in a future work, the kinetic barrier for H diffusion in HNO is very small (\~ {}0.1 eV), so H vacancies can readily hop between oxygen atoms.
Given structural relaxation, extra OH$^{-}$ ions adsorbed at nearby Nb$^{5+}$ sites move to OH vacancy sites on the top H$_2$Nb$_4$O$_8$ layer, and that effectively heals the Nb-OH-Nb link presented before surface dehydration. H vacancy below the surface plane is also healed by extra H ions from proton sources. In this way, HNO can recover its surface structure in low humidity environments. 

\begin{figure*}
\centering
\includegraphics[width=16cm]{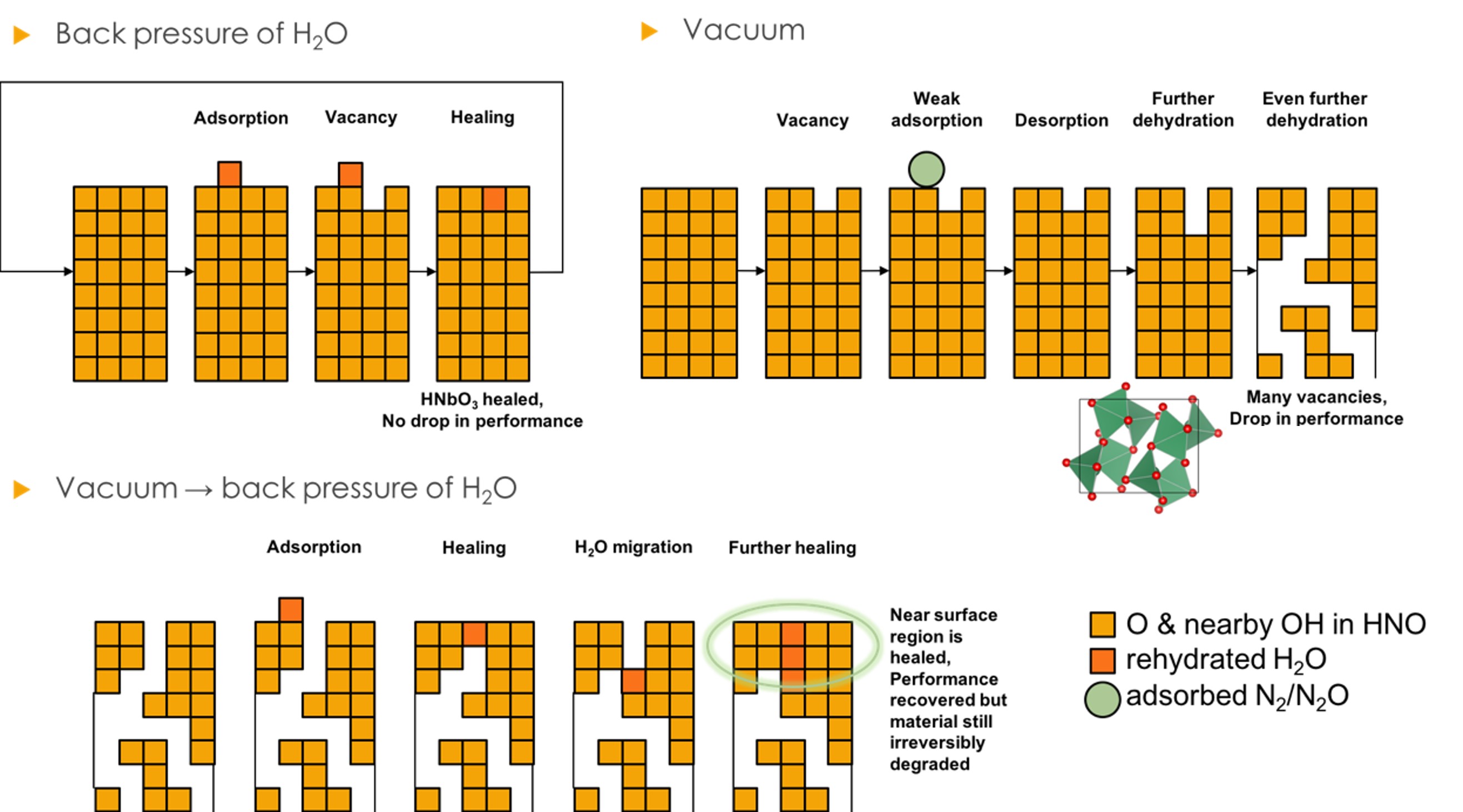}
\caption{\label{fig:illustration}
Illustration of the dehydration and rehydration process of HNO surface under different environmental conditions.} 
\end{figure*}

Based on these calculations, we propose the hypothesis for the optical performance of HNO-based materials under different environmental conditions (see Figure~\ref{fig:illustration}).
Under a back pressure of H$_2$O, the surface will be saturated with H$_2$O. If there are any OH or H vacancies, they will be immediately healed by the adsorbates.
If HNO is placed in vacuum, dehydration will occur, and that will cause a drastic change in local crystal structure and a downgrade in performance.
Finally, if the dehydrated surface is placed back in a humid environment, H$_2$O will be readsorbed on the surface and heal surface vacancy sites.

\section{Dielectric and optical properties of hydrogen niobate} 
\subsection{Low-frequency dielectric response of hydrogen niobate}
As the hydrogen niobate is the exchange product of lithium niobate, we first calculate and compare their dielectric properties. From linear response theory~\cite{Gonze97p10355}, the macroscopic low-frequency dielectric permittivity tensor $\epsilon_{\alpha\beta}(\omega)$ can be written as the sum of electronic ($\epsilon_{\alpha\beta}^{\infty}$) and ionic contributions: 
\begin{widetext}
\begin{align}
    \epsilon_{\alpha\beta}(\omega) = \epsilon_{\alpha\beta}^{\infty} + \frac{4\pi}{\Omega}\sum_{m} \frac{\bm{(}\sum_{\kappa\mu}Z^{*}_{\kappa,\alpha\mu}   U_{m,\bm{q}=0}(\kappa,\mu)\bm{)}\bm{(}\sum_{\kappa'\nu}Z^{*}_{\kappa',\beta\nu}U_{m,\bm{q}=0}(\kappa',\nu)\bm{)}}{\omega_m^2 - \omega^2},
\end{align}
\label{lowdielc}
\end{widetext}
where $\Omega$ is the volume of the unit cell,  $U_{m,q=0}(\kappa,\mu)$ is the eigen-displacement of phonon mode $m$ at wavevector $\bm{q} = 0$, and $Z^{*}_{\kappa,\alpha\mu}$ is the mode-effective charge vector. Subscripts $\alpha, \beta, \mu, \nu$ denote Cartesian coordinates, and $\kappa$ denotes atoms. The electronic contribution $\epsilon_{\alpha\beta}^{\infty}$ can be computed as:
\begin{equation}
    \epsilon_{\alpha\beta}^{\infty} = \delta_{\alpha\beta} - \frac{1}{\pi^2} \int_{BZ} \sum_{m}^{occ}\bra{\mu_{m\bm{k}}^{E_{\alpha}}}\ket{i\mu_{m\bm{k}}^{E_{\alpha}} \bm{k}},
\end{equation}
 where $\mu_{m\bm{k}}^{E_{\alpha}}$ is the first-order derivative of the Bloch wave functions with respect to an electric field along direction $\alpha$.
 Using density-functional perturbation theory~\cite{Gonze97p10355}, we first benchmark our method by computing static $(\omega = 0)$ dielectric tensor of rhombohedral LiNbO$_3$. As shown in Table I, our results are close to previously reported values, and the discrepancy originates from the different density functional we choose. This gives us confidence to compute the dielectric tensor of HNO.  
\begin{table}
\begin{tabular}{ |p{1.8cm}|p{1.8cm}|p{1.8cm}|p{1.8cm}|}
 \hline
  & $\epsilon^{0}_{xx}$ & $\epsilon^{0}_{yy}$ & $\epsilon^{0}_{zz}$ \\
 \hline
 Calc.~\cite{Veithen02p214302} & 42.4 & 42.4 & 29.3 \\
 \hline
 Exp.~\cite{Veithen02p214302} & 41.5 & 41.5 & 26.0 \\
 \hline
 This work & 36.5 & 36.5 & 28.7 \\ 
 \hline
\end{tabular}
 \caption{The static dielectric tensor of rhombohedral LiNbO$_3$.}
\end{table}

\begin{figure*}
\centering
\includegraphics[width=12 cm]{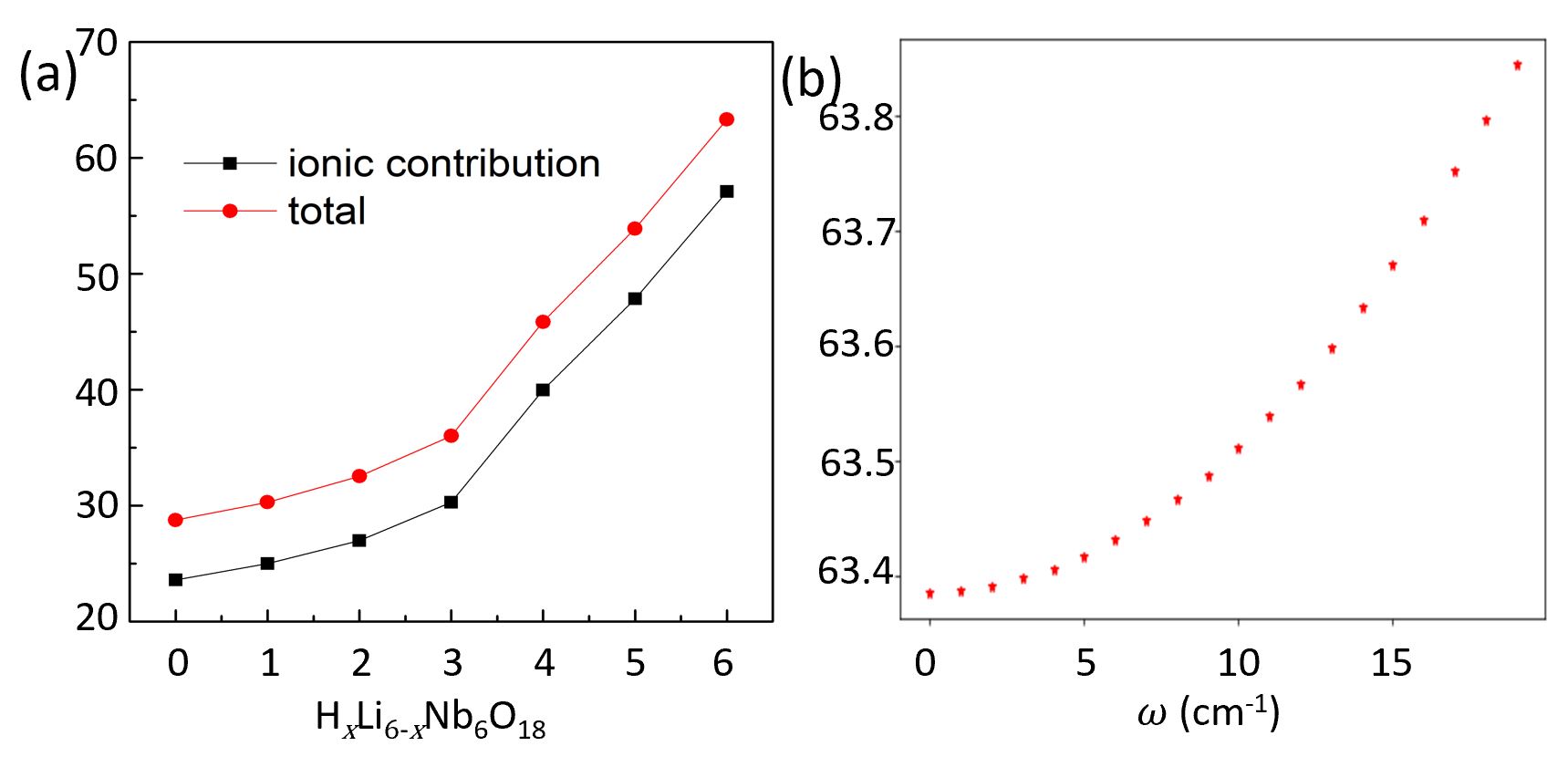}
\caption{\label{fig:LixH1-xdielec}
(a) Static dielectric $\epsilon_{zz}$ component of the H$_x$Li$_{6-x}$Nb$_{6}$O$_{18}$ with different $x$, constructed from the hexagonal conventional cell of  rhombohedral $\rm LiNbO_{3}$. (b) The dynamic dielectric $\epsilon_{zz}(\omega)$ component of the rhombohedral $\rm H_6 Nb_6 O_{18}$ with different frequencies of the electric field.}
\end{figure*}

We calculate the static dielectric tensor of compound H$_x$Li$_{6-x}$Nb$_{6}$O$_{18}$ by gradually replacing lithium ions with protons in the rhomobohedral phase of LiNbO$_3$.
As shown in Fig.~\ref{fig:LixH1-xdielec}(a), the total $\epsilon_{zz}$ is dramatically enhanced with the increasing $x$.
When all Li ions are replaced by protons in the rhombohedral LNO, static $\epsilon_{zz}$ is almost doubled.
We note the enhancement mainly comes from the increasingly dominant ionic contribution, while the electronic contribution is kept nearly constant over different proton concentrations.
This reveals that the dielectric response is heavily dependent on vibrational motions, dictated by compositions and structures of the system.
In Fig.~\ref{fig:LixH1-xdielec}(b) we show the dynamic low-frequency $\epsilon_{zz}(\omega)$ of the fully exchanged rhomobohedral H$_x$ Li$_{6-x}$Nb$_{6}$O$_{18}$ ($x = 6$). The dielectric tensor varies in a narrow window and is insensitive to the low-frequency driving field, as the field frequency is much smaller than the frequencies of the optical vibrational modes.   

\begin{figure}
\centering
\includegraphics[width=12 cm]{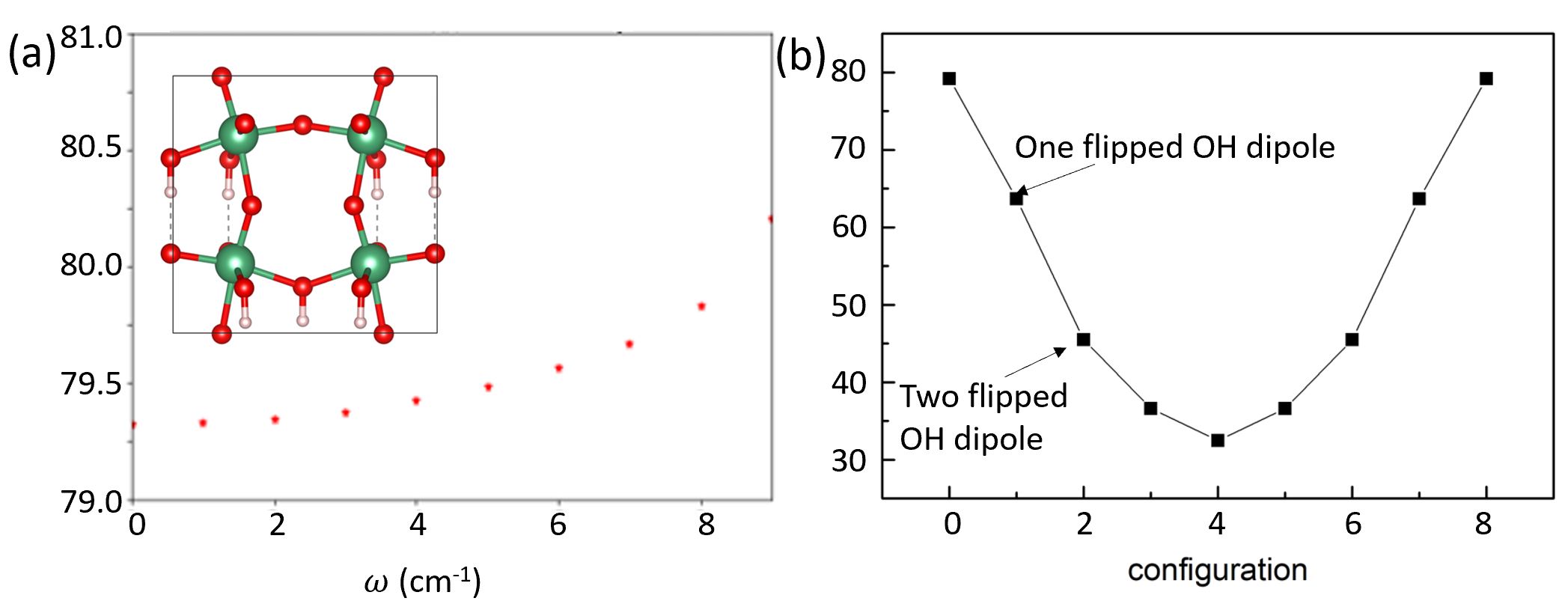}
\caption{\label{fig:dielectric}
(a) The dynamic dielectric $\epsilon_{zz}(\omega)$ component of the polar HNO with different frequencies of the electric field. (b) The static dielectric $\epsilon_{zz}(\omega = 0)$ with different numbers of flipped OH$^{-}$ dipoles.}
\end{figure}

Fig.~\ref{fig:dielectric}(a) shows the dielectric $\epsilon_{zz}(\omega)$ component of the low-frequency dielectric constant tensor of polar HNO. Similar to rhombohedral HNO (Fig.~\ref{fig:LixH1-xdielec}(b)), the total dielectric constant does not change much with the frequency of electric field. At $\omega =0$, $\epsilon_{zz}$ is around 79.25, almost 2.5 times larger than that of rhombohedral LNO. This can be possibly ascribed to the large polarization from eight polarized OH$^{-}$ dipoles.
To test that, Fig.~\ref{fig:dielectric}(b) shows the change of static dielectric constant by flipping OH$^{-}$ dipoles. The number on the horizontal axis represents the number of flipped OH$^{-}$ dipoles.
With more OH$^{-}$ flipped, the structure becomes less polarized, and it is fully unpolarized when four OH$^{-}$ dipoles are flipped.
Indeed, we see a dramatic decrease of $\epsilon_{zz}(0)$ with more flipped OH$^{-}$ dipoles, and the dielectric constant drops to its minimum when four OH$^{-}$ dipoles are flipped, only about 40$\%$ of the initial value.
After that, when more OH$^{-}$ dipoles are flipped, the structure becomes polarized again, though the polarization starts to rise along the opposite direction.
Fundamentally there is no difference between OH$^{-}$ dipoles pointing up or down; therefore we see a symmetric $\epsilon_{zz}(0)$ with the number of flipped OH$^{-}$ dipoles.

As we discuss in Section II, it is possible that both dehydrated and rehydrated HNO will form in ambient environment. In Table II, we list the diagonal components of the static dielectric tensors of HNO in all phases.
The off-diagonal components are much smaller, so they are not shown. The electronic contribution $\epsilon^{\infty}$ is very similar between these structures and much smaller than the ionic contribution.
The largest components are $\epsilon_{zz}$ of polar HNO (79.2) and $\epsilon_{xx}$ of rehydrated HNO (62.6). According to Eq. 5, frequencies of optical phonon modes $\omega_{m}$ are the decisive factors and they are smaller in polarized HNO, which result in a larger dielectric response. 

\begin{table}
\begin{tabular}{ |p{1cm}|p{1 cm}|p{1.4cm}|p{1.8cm}|p{1.8cm}|}
 \hline
   & LNO 
  & Polar HNO & Rehydrated HNO & Dehydrated HNO \\
 \hline
 $\epsilon_{xx}^{\infty}$ & 4.7 & 5.4 & 5.5 & 6.0\\
 \hline
 $\epsilon_{yy}^{\infty}$ & 4.7 & 5.4 & 5.2 & 5.6\\
 \hline
 $\epsilon_{zz}^{\infty}$ & 4.8 & 5.9 & 4.5 & 4.5\\
 \hline
 $\epsilon_{xx}$ & 36.5 & 56.4 & 62.6 & 34.5\\
 \hline
 $\epsilon_{yy}$ & 36.5 & 56.4 & 32.2 & 29.1\\
 \hline
 $\epsilon_{zz}$ & 28.7 & 79.2 & 25.2 & 10.1\\
 \hline
\end{tabular}
 \caption{Diagonal components of static dielectric tensors of rhombohedral LiNbO$_3$, polar, rehydrated and dehydrated HNO. Electronic contribution $\epsilon^{\infty}$ of each component is also given. }
\end{table}

\subsection{Refractive indices of hydrogen niobate}
To characterize the performance of HNO in waveguides, we calculate its refractive index as a function of light frequency.
The refractive index determines how much the path of light is bent or refracted when light propagates in a material.
Theoretically, it can be calculated through the imaginary and real part of frequency-dependent complex dielectric constant of the materials $\epsilon(\omega)$, 
\begin{gather}
    n_1 (\omega) = \sqrt{\frac{1}{2}\Big(\epsilon_1(\omega) + \sqrt{\epsilon^2_1(\omega) +\epsilon^2_2(\omega)}\Big)},\nonumber \\
    n_2 (\omega) = \frac{\epsilon^2_2(\omega)}{\sqrt{2\Big(\epsilon_1(\omega) + \sqrt{\epsilon^2_1(\omega) + \epsilon^2_2(\omega)}\Big)}},
\end{gather}
where $n_1(\omega)$ and $n_2(\omega)$, $\epsilon_1(\omega)$ and $\epsilon_2(\omega)$ are the real and imaginary parts of the refractive index and the complex dielectric constant, respectively.
When $\hbar\omega$ falls in the range of visible spectrum, $\epsilon(\omega)$ has a different form from Eq.5. From linear response theory, the imaginary part $\epsilon_2(\omega)$ can be calculated as~\cite{Rohlfing00p4927}
\begin{equation}
    \epsilon_2(\omega) = \frac{16\pi e^2}{\omega^2} \sum_{c,v} |\bra{v}\Vec{v}\ket{c}|^{2}\delta[\hbar\omega-(E_c - E_v)],\label{opticalexcitation}
\end{equation}
where $\bra{v}\Vec{v}\ket{c}$ is the dipole matrix element between conduction band $c$ and valence band $v$. With $\epsilon_2(\omega)$, the real part $\epsilon_1(\omega)$ can be computed via Kramers-Kr$\rm \Ddot{o}$nig relation
\begin{equation}
     \epsilon_1(\omega) = 1 + \frac{2}{\pi}\mathcal{P}\int_{0}^{\infty} d\omega' \frac{\omega'\epsilon_2(\omega)}{\omega'^2 - \omega^2}.
\end{equation}.

\begin{figure}
\centering
\includegraphics[width=12 cm]{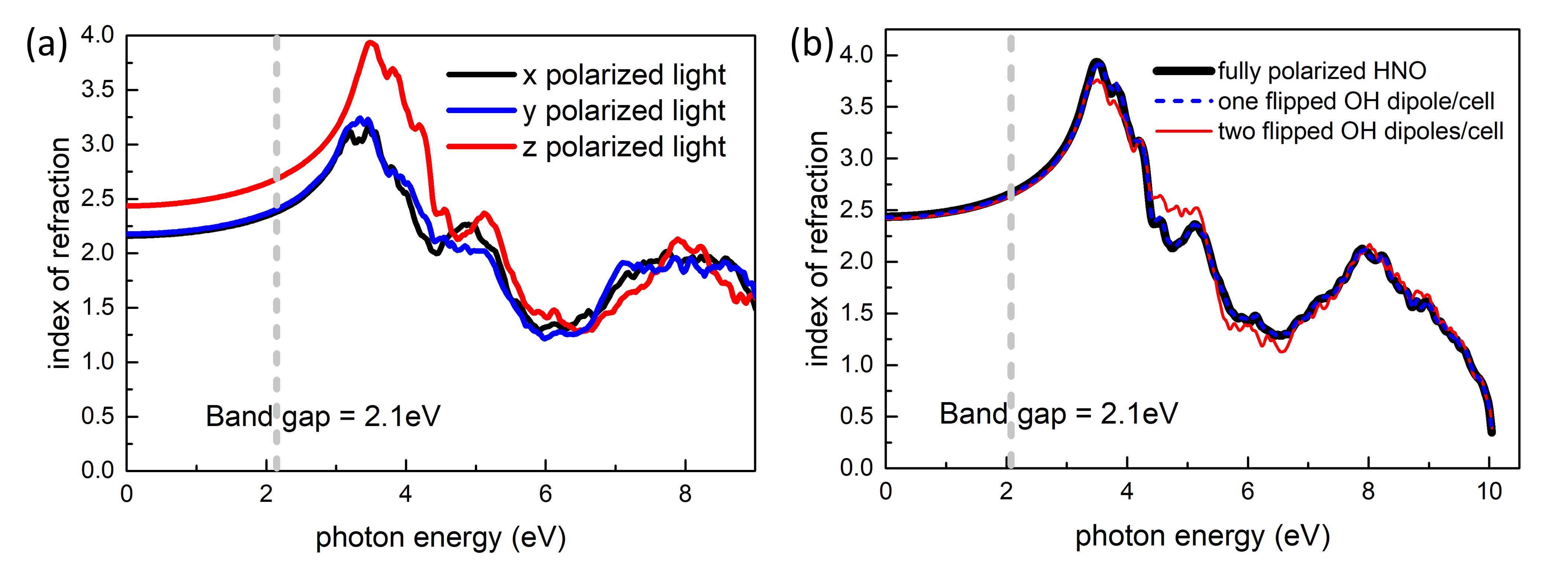}
\caption{\label{fig:polarindex}
(a) Refractive index of polar HNO with different light polarizations. (b) Refractive indices of HNO with different numbers of flipped OH$^{-}$ dipoles.} 
\end{figure}

We first examine the effect of the polarization of light on the refractive index. The polarization direction determines which component of the dipole matrix element is coupled with light.
As shown in Fig.~\ref{fig:polarindex}(a), for polar HNO, when the photon energy is smaller than the band gap, the refractive index for $z-$polarized light ($n_{xx}\approx$2.4) is close to indices of $x-$ and $y-$polarized light ($n_{xx/yy}\approx$2.2).
When $\hbar\omega$ is larger than band gap (2.1 eV), with resonant excitation and optical transitions across the band gap, there is a sharp increase of refractive index for $n_{zz}$, and it reaches 4.0 when $\hbar\omega$ = 3.7 eV.
The increase of $n_{zz}$ is more significant than $n_{xx}$ and $n_{yy}$, consistent with a larger dielectric response $\epsilon^{\infty}_{zz}$ with purely electronic origin. 
As a comparison, in LNO, for below-band-gap excitation, the refractive index stays around 2.2, and then it increases to 2.44 when $\hbar\omega = 3.1$ eV~\cite{Zelmon97p3319}.
Therefore, the contrast of refractive indices at the HNbO$_3$/LiNbO$_3$ interface in waveguides will be strong at shorter wavelengths.   

To study the relationship between optical response and macroscopic polarization, we plot refractive indices over the number of flipped OH$^{-}$ dipoles. As shown in Fig.~\ref{fig:polarindex}(b), the refractive index does not change much with the orientation of OH$^{-}$ dipoles.
This is because optical excitation characterizes transition between different electronic bands, more correlated with dielectric response from electronic contribution rather than from ionic contribution.
As shown in Table II, the ionic part differs more than the electronic part between different phases of HNO.
As such, we expect the refractive indices are similar for these OH$^{-}$-flipped structures.

\begin{figure}
\centering
\includegraphics[width=8 cm]{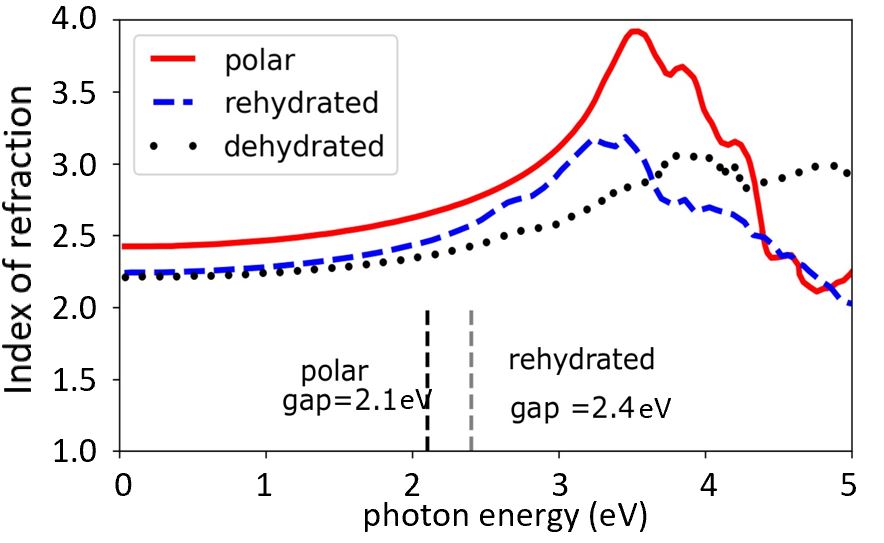}
\caption{\label{fig:refractiveindex}
The absolute value of refractive indices
$\epsilon_{zz}$ of polar, rehydrated, and
dehydrated HNO.
} 
\end{figure}

Figure~\ref{fig:refractiveindex}  shows the refractive indices for polar, rehydrated and dehydrated HNOs. In general, they all exhibit a similar tendency with the change of photon energy.
Below band gap, $\epsilon_{zz}$ of polar HNO is 0.2 higher than that of rehydrated and dehydrated HNO. Above band gap, $\epsilon_{zz}$ is much larger in polar HNO (4.0) than in the other two phases (3.0), and they peak at different photon energies. In dehydrated HNO, peak frequencies of $n_{xx}$, $n_{yy}$ and $n_{zz}$ are also not aligned and vary between 3-4 eV (not shown here). As a consequence, when light propagates in areas where all three phases coexist or get mixed, the propagation will be inhomogenous due to the phase mismatch between light beams. 

\section{Conclusion}
In conclusion, by performing first principles calculations, we study the ground state structure of bulk hydrogen niobate and related surfaces. Thermodynamics characterize the dehydration and rehydration processes of polar hydrogen niobate under different ambient conditions. We further study its low-frequency dielectric response and optical refractive indices in different phases. Our study shows that hydrogen niobate exhibits excellent chemical and physical properties, and is promising for the future application of photonic devices.

\section{Acknowledgements}
{This work was supported by the Office of Naval Research, under grant N00014-20-1-2701, and by the center for 3D Ferroelectric Microelectronics (3DFeM), an Energy Frontier Research Center funded by the U.S. Department of Energy (DOE), Office of Science, Basic Energy Sciences under Award No.DE-SC0021118. Computational support was provided by the High Performance Computing Modernization Program of the US Department of Defense, and by the National Energy Research Scientific Computing Center (NERSC), a U.S. Department of Energy, Office of Science User Facility located at Lawrence Berkeley National Laboratory, operated under Contract No. DE-AC02-05CH11231..}       
\bibliography{rappecites.bib}

\begin{thebibliography}{39}%
\makeatletter
\providecommand \@ifxundefined [1]{%
 \@ifx{#1\undefined}
}%
\providecommand \@ifnum [1]{%
 \ifnum #1\expandafter \@firstoftwo
 \else \expandafter \@secondoftwo
 \fi
}%
\providecommand \@ifx [1]{%
 \ifx #1\expandafter \@firstoftwo
 \else \expandafter \@secondoftwo
 \fi
}%
\providecommand \natexlab [1]{#1}%
\providecommand \enquote  [1]{``#1''}%
\providecommand \bibnamefont  [1]{#1}%
\providecommand \bibfnamefont [1]{#1}%
\providecommand \citenamefont [1]{#1}%
\providecommand \href@noop [0]{\@secondoftwo}%
\providecommand \href [0]{\begingroup \@sanitize@url \@href}%
\providecommand \@href[1]{\@@startlink{#1}\@@href}%
\providecommand \@@href[1]{\endgroup#1\@@endlink}%
\providecommand \@sanitize@url [0]{\catcode `\\12\catcode `\$12\catcode
  `\&12\catcode `\#12\catcode `\^12\catcode `\_12\catcode `\%12\relax}%
\providecommand \@@startlink[1]{}%
\providecommand \@@endlink[0]{}%
\providecommand \url  [0]{\begingroup\@sanitize@url \@url }%
\providecommand \@url [1]{\endgroup\@href {#1}{\urlprefix }}%
\providecommand \urlprefix  [0]{URL }%
\providecommand \Eprint [0]{\href }%
\providecommand \doibase [0]{http://dx.doi.org/}%
\providecommand \selectlanguage [0]{\@gobble}%
\providecommand \bibinfo  [0]{\@secondoftwo}%
\providecommand \bibfield  [0]{\@secondoftwo}%
\providecommand \translation [1]{[#1]}%
\providecommand \BibitemOpen [0]{}%
\providecommand \bibitemStop [0]{}%
\providecommand \bibitemNoStop [0]{.\EOS\space}%
\providecommand \EOS [0]{\spacefactor3000\relax}%
\providecommand \BibitemShut  [1]{\csname bibitem#1\endcsname}%
\let\auto@bib@innerbib\@empty
\bibitem [{\citenamefont {Carruthers}\ \emph {et~al.}(1971)\citenamefont
  {Carruthers}, \citenamefont {Peterson},\ and\ \citenamefont
  {Grasso}}]{Carruthers71p1846}%
  \BibitemOpen
  \bibfield  {author} {\bibinfo {author} {\bibfnamefont {J.}~\bibnamefont
  {Carruthers}}, \bibinfo {author} {\bibfnamefont {G.}~\bibnamefont
  {Peterson}}, \ and\ \bibinfo {author} {\bibfnamefont {M.}~\bibnamefont
  {Grasso}},\ }\href@noop {} {\bibfield  {journal} {\bibinfo  {journal} {J.
  App. Phys.}\ }\textbf {\bibinfo {volume} {42}},\ \bibinfo {pages} {1846}
  (\bibinfo {year} {1971})}\BibitemShut {NoStop}%
\bibitem [{\citenamefont {Nishihara}\ \emph {et~al.}(2000)\citenamefont
  {Nishihara}, \citenamefont {Haruna},\ and\ \citenamefont
  {Suhara}}]{nishihara00p26}%
  \BibitemOpen
  \bibfield  {author} {\bibinfo {author} {\bibfnamefont {H.}~\bibnamefont
  {Nishihara}}, \bibinfo {author} {\bibfnamefont {M.}~\bibnamefont {Haruna}}, \
  and\ \bibinfo {author} {\bibfnamefont {T.}~\bibnamefont {Suhara}},\
  }\href@noop {} {\bibfield  {journal} {\bibinfo  {journal} {Electro-Optics
  Handbook}\ } (\bibinfo {year} {2000})}\BibitemShut {NoStop}%
\bibitem [{\citenamefont {Wooten}\ \emph {et~al.}(2000)\citenamefont {Wooten},
  \citenamefont {Kissa}, \citenamefont {Yi-Yan}, \citenamefont {Murphy},
  \citenamefont {Lafaw}, \citenamefont {Hallemeier}, \citenamefont {Maack},
  \citenamefont {Attanasio}, \citenamefont {Fritz}, \citenamefont {McBrien}
  \emph {et~al.}}]{Wooten00p69}%
  \BibitemOpen
  \bibfield  {author} {\bibinfo {author} {\bibfnamefont {E.~L.}\ \bibnamefont
  {Wooten}}, \bibinfo {author} {\bibfnamefont {K.~M.}\ \bibnamefont {Kissa}},
  \bibinfo {author} {\bibfnamefont {A.}~\bibnamefont {Yi-Yan}}, \bibinfo
  {author} {\bibfnamefont {E.~J.}\ \bibnamefont {Murphy}}, \bibinfo {author}
  {\bibfnamefont {D.~A.}\ \bibnamefont {Lafaw}}, \bibinfo {author}
  {\bibfnamefont {P.~F.}\ \bibnamefont {Hallemeier}}, \bibinfo {author}
  {\bibfnamefont {D.}~\bibnamefont {Maack}}, \bibinfo {author} {\bibfnamefont
  {D.~V.}\ \bibnamefont {Attanasio}}, \bibinfo {author} {\bibfnamefont {D.~J.}\
  \bibnamefont {Fritz}}, \bibinfo {author} {\bibfnamefont {G.~J.}\ \bibnamefont
  {McBrien}},  \emph {et~al.},\ }\href@noop {} {\bibfield  {journal} {\bibinfo
  {journal} {IEEE Journal of selected topics in Quantum Electronics}\ }\textbf
  {\bibinfo {volume} {6}},\ \bibinfo {pages} {69} (\bibinfo {year}
  {2000})}\BibitemShut {NoStop}%
\bibitem [{\citenamefont {Chen}\ \emph {et~al.}(1995)\citenamefont {Chen},
  \citenamefont {LaMacchia},\ and\ \citenamefont {Fraser}}]{Chen95p33}%
  \BibitemOpen
  \bibfield  {author} {\bibinfo {author} {\bibfnamefont {F.}~\bibnamefont
  {Chen}}, \bibinfo {author} {\bibfnamefont {J.}~\bibnamefont {LaMacchia}}, \
  and\ \bibinfo {author} {\bibfnamefont {D.}~\bibnamefont {Fraser}},\ }in\
  \href@noop {} {\emph {\bibinfo {booktitle} {Landmark Papers on
  Photorefractive Nonlinear Optics}}}\ (\bibinfo  {publisher} {World
  Scientific},\ \bibinfo {year} {1995})\ pp.\ \bibinfo {pages}
  {33--35}\BibitemShut {NoStop}%
\bibitem [{\citenamefont {de~Miguel-Sanz}\ \emph {et~al.}(2002)\citenamefont
  {de~Miguel-Sanz}, \citenamefont {Carrascosa},\ and\ \citenamefont
  {Arizmendi}}]{Miguel-Sanz65p165101}%
  \BibitemOpen
  \bibfield  {author} {\bibinfo {author} {\bibfnamefont {E.}~\bibnamefont
  {de~Miguel-Sanz}}, \bibinfo {author} {\bibfnamefont {M.}~\bibnamefont
  {Carrascosa}}, \ and\ \bibinfo {author} {\bibfnamefont {L.}~\bibnamefont
  {Arizmendi}},\ }\href@noop {} {\bibfield  {journal} {\bibinfo  {journal}
  {Physical Review B}\ }\textbf {\bibinfo {volume} {65}},\ \bibinfo {pages}
  {165101} (\bibinfo {year} {2002})}\BibitemShut {NoStop}%
\bibitem [{\citenamefont {Zhu}\ and\ \citenamefont {Ming}(1992)}]{Zhu92p904}%
  \BibitemOpen
  \bibfield  {author} {\bibinfo {author} {\bibfnamefont {Y.-y.}\ \bibnamefont
  {Zhu}}\ and\ \bibinfo {author} {\bibfnamefont {N.-b.}\ \bibnamefont {Ming}},\
  }\href@noop {} {\bibfield  {journal} {\bibinfo  {journal} {Journal of applied
  physics}\ }\textbf {\bibinfo {volume} {72}},\ \bibinfo {pages} {904}
  (\bibinfo {year} {1992})}\BibitemShut {NoStop}%
\bibitem [{\citenamefont {Pruneri}\ \emph {et~al.}(1995)\citenamefont
  {Pruneri}, \citenamefont {Koch}, \citenamefont {Kazansky}, \citenamefont
  {Clarkson}, \citenamefont {Russell},\ and\ \citenamefont
  {Hanna}}]{Pruneri95p2375}%
  \BibitemOpen
  \bibfield  {author} {\bibinfo {author} {\bibfnamefont {V.}~\bibnamefont
  {Pruneri}}, \bibinfo {author} {\bibfnamefont {R.}~\bibnamefont {Koch}},
  \bibinfo {author} {\bibfnamefont {P.}~\bibnamefont {Kazansky}}, \bibinfo
  {author} {\bibfnamefont {W.}~\bibnamefont {Clarkson}}, \bibinfo {author}
  {\bibfnamefont {P.~S.~J.}\ \bibnamefont {Russell}}, \ and\ \bibinfo {author}
  {\bibfnamefont {D.}~\bibnamefont {Hanna}},\ }\href@noop {} {\bibfield
  {journal} {\bibinfo  {journal} {Optics letters}\ }\textbf {\bibinfo {volume}
  {20}},\ \bibinfo {pages} {2375} (\bibinfo {year} {1995})}\BibitemShut
  {NoStop}%
\bibitem [{\citenamefont {K{\"o}sters}\ \emph {et~al.}(2009)\citenamefont
  {K{\"o}sters}, \citenamefont {Sturman}, \citenamefont {Werheit},
  \citenamefont {Haertle},\ and\ \citenamefont {Buse}}]{Kosters09p510}%
  \BibitemOpen
  \bibfield  {author} {\bibinfo {author} {\bibfnamefont {M.}~\bibnamefont
  {K{\"o}sters}}, \bibinfo {author} {\bibfnamefont {B.}~\bibnamefont
  {Sturman}}, \bibinfo {author} {\bibfnamefont {P.}~\bibnamefont {Werheit}},
  \bibinfo {author} {\bibfnamefont {D.}~\bibnamefont {Haertle}}, \ and\
  \bibinfo {author} {\bibfnamefont {K.}~\bibnamefont {Buse}},\ }\href@noop {}
  {\bibfield  {journal} {\bibinfo  {journal} {Nature Photonics}\ }\textbf
  {\bibinfo {volume} {3}},\ \bibinfo {pages} {510} (\bibinfo {year}
  {2009})}\BibitemShut {NoStop}%
\bibitem [{\citenamefont {Schmidt}\ and\ \citenamefont
  {Kaminow}(1974)}]{Schmidt74p458}%
  \BibitemOpen
  \bibfield  {author} {\bibinfo {author} {\bibfnamefont {R.}~\bibnamefont
  {Schmidt}}\ and\ \bibinfo {author} {\bibfnamefont {I.}~\bibnamefont
  {Kaminow}},\ }\href@noop {} {\bibfield  {journal} {\bibinfo  {journal}
  {Applied Physics Letters}\ }\textbf {\bibinfo {volume} {25}},\ \bibinfo
  {pages} {458} (\bibinfo {year} {1974})}\BibitemShut {NoStop}%
\bibitem [{\citenamefont {Arizmendi}(2004)}]{arizmendi2004photonic}%
  \BibitemOpen
  \bibfield  {author} {\bibinfo {author} {\bibfnamefont {L.}~\bibnamefont
  {Arizmendi}},\ }\href@noop {} {\bibfield  {journal} {\bibinfo  {journal}
  {physica status solidi (a)}\ }\textbf {\bibinfo {volume} {201}},\ \bibinfo
  {pages} {253} (\bibinfo {year} {2004})}\BibitemShut {NoStop}%
\bibitem [{\citenamefont {Neyer}\ and\ \citenamefont
  {Sohler}(1979)}]{Neyer79p256}%
  \BibitemOpen
  \bibfield  {author} {\bibinfo {author} {\bibfnamefont {A.}~\bibnamefont
  {Neyer}}\ and\ \bibinfo {author} {\bibfnamefont {W.}~\bibnamefont {Sohler}},\
  }\href@noop {} {\bibfield  {journal} {\bibinfo  {journal} {Applied Physics
  Letters}\ }\textbf {\bibinfo {volume} {35}},\ \bibinfo {pages} {256}
  (\bibinfo {year} {1979})}\BibitemShut {NoStop}%
\bibitem [{\citenamefont {Fukuma}\ and\ \citenamefont
  {Noda}(1980)}]{Fukuma80p591}%
  \BibitemOpen
  \bibfield  {author} {\bibinfo {author} {\bibfnamefont {M.}~\bibnamefont
  {Fukuma}}\ and\ \bibinfo {author} {\bibfnamefont {J.}~\bibnamefont {Noda}},\
  }\href@noop {} {\bibfield  {journal} {\bibinfo  {journal} {Applied optics}\
  }\textbf {\bibinfo {volume} {19}},\ \bibinfo {pages} {591} (\bibinfo {year}
  {1980})}\BibitemShut {NoStop}%
\bibitem [{\citenamefont {Jackel}\ \emph {et~al.}(1982)\citenamefont {Jackel},
  \citenamefont {Rice},\ and\ \citenamefont {Veselka}}]{Jackel82p607}%
  \BibitemOpen
  \bibfield  {author} {\bibinfo {author} {\bibfnamefont {J.~L.}\ \bibnamefont
  {Jackel}}, \bibinfo {author} {\bibfnamefont {C.}~\bibnamefont {Rice}}, \ and\
  \bibinfo {author} {\bibfnamefont {J.}~\bibnamefont {Veselka}},\ }\href@noop
  {} {\bibfield  {journal} {\bibinfo  {journal} {Applied Physics Letters}\
  }\textbf {\bibinfo {volume} {41}},\ \bibinfo {pages} {607} (\bibinfo {year}
  {1982})}\BibitemShut {NoStop}%
\bibitem [{\citenamefont {Cabrera}\ \emph {et~al.}(1996)\citenamefont
  {Cabrera}, \citenamefont {Olivares}, \citenamefont {Carrascosa},
  \citenamefont {Rams}, \citenamefont {M{\"u}ller},\ and\ \citenamefont
  {Di{\'e}guez}}]{Cabrera96p349}%
  \BibitemOpen
  \bibfield  {author} {\bibinfo {author} {\bibfnamefont {J.}~\bibnamefont
  {Cabrera}}, \bibinfo {author} {\bibfnamefont {J.}~\bibnamefont {Olivares}},
  \bibinfo {author} {\bibfnamefont {M.}~\bibnamefont {Carrascosa}}, \bibinfo
  {author} {\bibfnamefont {J.}~\bibnamefont {Rams}}, \bibinfo {author}
  {\bibfnamefont {R.}~\bibnamefont {M{\"u}ller}}, \ and\ \bibinfo {author}
  {\bibfnamefont {E.}~\bibnamefont {Di{\'e}guez}},\ }\href@noop {} {\bibfield
  {journal} {\bibinfo  {journal} {Advances in Physics}\ }\textbf {\bibinfo
  {volume} {45}},\ \bibinfo {pages} {349} (\bibinfo {year} {1996})}\BibitemShut
  {NoStop}%
\bibitem [{\citenamefont {Rice}(1986)}]{Rice86p188}%
  \BibitemOpen
  \bibfield  {author} {\bibinfo {author} {\bibfnamefont {C.}~\bibnamefont
  {Rice}},\ }\href@noop {} {\bibfield  {journal} {\bibinfo  {journal} {Journal
  of Solid State Chemistry}\ }\textbf {\bibinfo {volume} {64}},\ \bibinfo
  {pages} {188} (\bibinfo {year} {1986})}\BibitemShut {NoStop}%
\bibitem [{\citenamefont {Korkishko}\ \emph {et~al.}(1996)\citenamefont
  {Korkishko}, \citenamefont {Fedorov}, \citenamefont {De~Micheli},
  \citenamefont {Baldi}, \citenamefont {El~Hadi},\ and\ \citenamefont
  {Leycuras}}]{korkishko96p7056}%
  \BibitemOpen
  \bibfield  {author} {\bibinfo {author} {\bibfnamefont {Y.~N.}\ \bibnamefont
  {Korkishko}}, \bibinfo {author} {\bibfnamefont {V.}~\bibnamefont {Fedorov}},
  \bibinfo {author} {\bibfnamefont {M.}~\bibnamefont {De~Micheli}}, \bibinfo
  {author} {\bibfnamefont {P.}~\bibnamefont {Baldi}}, \bibinfo {author}
  {\bibfnamefont {K.}~\bibnamefont {El~Hadi}}, \ and\ \bibinfo {author}
  {\bibfnamefont {A.}~\bibnamefont {Leycuras}},\ }\href@noop {} {\bibfield
  {journal} {\bibinfo  {journal} {Applied optics}\ }\textbf {\bibinfo {volume}
  {35}},\ \bibinfo {pages} {7056} (\bibinfo {year} {1996})}\BibitemShut
  {NoStop}%
\bibitem [{\citenamefont {Korkishko}\ and\ \citenamefont
  {Fedorov}(1996{\natexlab{a}})}]{Korkishko96p175}%
  \BibitemOpen
  \bibfield  {author} {\bibinfo {author} {\bibfnamefont {Y.~N.}\ \bibnamefont
  {Korkishko}}\ and\ \bibinfo {author} {\bibfnamefont {V.}~\bibnamefont
  {Fedorov}},\ }\href@noop {} {\bibfield  {journal} {\bibinfo  {journal}
  {Optical Materials}\ }\textbf {\bibinfo {volume} {5}},\ \bibinfo {pages}
  {175} (\bibinfo {year} {1996}{\natexlab{a}})}\BibitemShut {NoStop}%
\bibitem [{\citenamefont {Korkishko}\ \emph {et~al.}(1995)\citenamefont
  {Korkishko}, \citenamefont {Fedorov}, \citenamefont {Katin},\ and\
  \citenamefont {Kondrat'ev}}]{korkishko95p149}%
  \BibitemOpen
  \bibfield  {author} {\bibinfo {author} {\bibfnamefont {Y.~N.}\ \bibnamefont
  {Korkishko}}, \bibinfo {author} {\bibfnamefont {V.~A.}\ \bibnamefont
  {Fedorov}}, \bibinfo {author} {\bibfnamefont {S.}~\bibnamefont {Katin}}, \
  and\ \bibinfo {author} {\bibfnamefont {A.~V.}\ \bibnamefont {Kondrat'ev}},\
  }in\ \href@noop {} {\emph {\bibinfo {booktitle} {Functional Photonic
  Integrated Circuits}}},\ Vol.\ \bibinfo {volume} {2401}\ (\bibinfo
  {organization} {SPIE},\ \bibinfo {year} {1995})\ pp.\ \bibinfo {pages}
  {149--161}\BibitemShut {NoStop}%
\bibitem [{\citenamefont {Korkishko}\ and\ \citenamefont
  {Fedorov}(1996{\natexlab{b}})}]{Korkishko96p187}%
  \BibitemOpen
  \bibfield  {author} {\bibinfo {author} {\bibfnamefont {Y.~N.}\ \bibnamefont
  {Korkishko}}\ and\ \bibinfo {author} {\bibfnamefont {V.}~\bibnamefont
  {Fedorov}},\ }\href@noop {} {\bibfield  {journal} {\bibinfo  {journal} {IEEE
  Journal of Selected Topics in Quantum Electronics}\ }\textbf {\bibinfo
  {volume} {2}},\ \bibinfo {pages} {187} (\bibinfo {year}
  {1996}{\natexlab{b}})}\BibitemShut {NoStop}%
\bibitem [{\citenamefont {Korkishko}\ \emph {et~al.}(1998)\citenamefont
  {Korkishko}, \citenamefont {Fedorov}, \citenamefont {Morozova}, \citenamefont
  {Caccavale}, \citenamefont {Gonella},\ and\ \citenamefont
  {Segato}}]{korkishko98p1838}%
  \BibitemOpen
  \bibfield  {author} {\bibinfo {author} {\bibfnamefont {Y.~N.}\ \bibnamefont
  {Korkishko}}, \bibinfo {author} {\bibfnamefont {V.}~\bibnamefont {Fedorov}},
  \bibinfo {author} {\bibfnamefont {T.}~\bibnamefont {Morozova}}, \bibinfo
  {author} {\bibfnamefont {F.}~\bibnamefont {Caccavale}}, \bibinfo {author}
  {\bibfnamefont {F.}~\bibnamefont {Gonella}}, \ and\ \bibinfo {author}
  {\bibfnamefont {F.}~\bibnamefont {Segato}},\ }\href@noop {} {\bibfield
  {journal} {\bibinfo  {journal} {JOSA A}\ }\textbf {\bibinfo {volume} {15}},\
  \bibinfo {pages} {1838} (\bibinfo {year} {1998})}\BibitemShut {NoStop}%
\bibitem [{\citenamefont {Ohsaka}\ \emph {et~al.}(2001)\citenamefont {Ohsaka},
  \citenamefont {Kanzaki},\ and\ \citenamefont {Abe}}]{Ohsaka01p2141}%
  \BibitemOpen
  \bibfield  {author} {\bibinfo {author} {\bibfnamefont {T.}~\bibnamefont
  {Ohsaka}}, \bibinfo {author} {\bibfnamefont {Y.}~\bibnamefont {Kanzaki}}, \
  and\ \bibinfo {author} {\bibfnamefont {M.}~\bibnamefont {Abe}},\ }\href@noop
  {} {\bibfield  {journal} {\bibinfo  {journal} {Materials research bulletin}\
  }\textbf {\bibinfo {volume} {36}},\ \bibinfo {pages} {2141} (\bibinfo {year}
  {2001})}\BibitemShut {NoStop}%
\bibitem [{\citenamefont {Fedorov}\ and\ \citenamefont
  {Korkishko}(1994)}]{Fedorov94p243}%
  \BibitemOpen
  \bibfield  {author} {\bibinfo {author} {\bibfnamefont {V.~A.}\ \bibnamefont
  {Fedorov}}\ and\ \bibinfo {author} {\bibfnamefont {Y.~N.}\ \bibnamefont
  {Korkishko}},\ }in\ \href@noop {} {\emph {\bibinfo {booktitle} {Integrated
  Optics and Microstructures II}}},\ Vol.\ \bibinfo {volume} {2291}\ (\bibinfo
  {organization} {International Society for Optics and Photonics},\ \bibinfo
  {year} {1994})\ pp.\ \bibinfo {pages} {243--255}\BibitemShut {NoStop}%
\bibitem [{\citenamefont {Fedorov}\ and\ \citenamefont
  {Korkishko}(1995)}]{fedorov95p216}%
  \BibitemOpen
  \bibfield  {author} {\bibinfo {author} {\bibfnamefont {V.~A.}\ \bibnamefont
  {Fedorov}}\ and\ \bibinfo {author} {\bibfnamefont {Y.~N.}\ \bibnamefont
  {Korkishko}},\ }in\ \href@noop {} {\emph {\bibinfo {booktitle} {Functional
  Photonic Integrated Circuits}}},\ Vol.\ \bibinfo {volume} {2401}\ (\bibinfo
  {organization} {International Society for Optics and Photonics},\ \bibinfo
  {year} {1995})\ pp.\ \bibinfo {pages} {216--226}\BibitemShut {NoStop}%
\bibitem [{\citenamefont {Korkishko}\ \emph {et~al.}(1997)\citenamefont
  {Korkishko}, \citenamefont {Fedorov}, \citenamefont {Nosikov}, \citenamefont
  {Kostritskii},\ and\ \citenamefont {De~Micheli}}]{Korkishko97p188}%
  \BibitemOpen
  \bibfield  {author} {\bibinfo {author} {\bibfnamefont {Y.~N.}\ \bibnamefont
  {Korkishko}}, \bibinfo {author} {\bibfnamefont {V.~A.}\ \bibnamefont
  {Fedorov}}, \bibinfo {author} {\bibfnamefont {V.~V.}\ \bibnamefont
  {Nosikov}}, \bibinfo {author} {\bibfnamefont {S.~M.}\ \bibnamefont
  {Kostritskii}}, \ and\ \bibinfo {author} {\bibfnamefont {M.~P.}\ \bibnamefont
  {De~Micheli}},\ }in\ \href@noop {} {\emph {\bibinfo {booktitle} {Integrated
  Optics Devices: Potential for Commercialization}}},\ Vol.\ \bibinfo {volume}
  {2997}\ (\bibinfo {organization} {International Society for Optics and
  Photonics},\ \bibinfo {year} {1997})\ pp.\ \bibinfo {pages}
  {188--200}\BibitemShut {NoStop}%
\bibitem [{\citenamefont {Rice}\ and\ \citenamefont
  {Jackel}(1982)}]{Rice82p308}%
  \BibitemOpen
  \bibfield  {author} {\bibinfo {author} {\bibfnamefont {C.}~\bibnamefont
  {Rice}}\ and\ \bibinfo {author} {\bibfnamefont {J.}~\bibnamefont {Jackel}},\
  }\href@noop {} {\bibfield  {journal} {\bibinfo  {journal} {Journal of Solid
  State Chemistry}\ }\textbf {\bibinfo {volume} {41}},\ \bibinfo {pages} {308}
  (\bibinfo {year} {1982})}\BibitemShut {NoStop}%
\bibitem [{\citenamefont {Fourquet}\ \emph {et~al.}(1983)\citenamefont
  {Fourquet}, \citenamefont {Renou}, \citenamefont {De~Pape}, \citenamefont
  {Theveneau}, \citenamefont {Man}, \citenamefont {Lucas},\ and\ \citenamefont
  {Pannetier}}]{Fourquet83p1011}%
  \BibitemOpen
  \bibfield  {author} {\bibinfo {author} {\bibfnamefont {J.}~\bibnamefont
  {Fourquet}}, \bibinfo {author} {\bibfnamefont {M.}~\bibnamefont {Renou}},
  \bibinfo {author} {\bibfnamefont {R.}~\bibnamefont {De~Pape}}, \bibinfo
  {author} {\bibfnamefont {H.}~\bibnamefont {Theveneau}}, \bibinfo {author}
  {\bibfnamefont {P.}~\bibnamefont {Man}}, \bibinfo {author} {\bibfnamefont
  {O.}~\bibnamefont {Lucas}}, \ and\ \bibinfo {author} {\bibfnamefont
  {J.}~\bibnamefont {Pannetier}},\ }\href@noop {} {\bibfield  {journal}
  {\bibinfo  {journal} {Solid State Ionics}\ }\textbf {\bibinfo {volume} {9}},\
  \bibinfo {pages} {1011} (\bibinfo {year} {1983})}\BibitemShut {NoStop}%
\bibitem [{\citenamefont {Pokrovskii}(2000)}]{Pokrovskii00P890}%
  \BibitemOpen
  \bibfield  {author} {\bibinfo {author} {\bibfnamefont {L.}~\bibnamefont
  {Pokrovskii}},\ }\href@noop {} {\bibfield  {journal} {\bibinfo  {journal}
  {Journal of Structural Chemistry}\ }\textbf {\bibinfo {volume} {41}},\
  \bibinfo {pages} {890} (\bibinfo {year} {2000})}\BibitemShut {NoStop}%
\bibitem [{\citenamefont {Kalabin}\ \emph {et~al.}(2003)\citenamefont
  {Kalabin}, \citenamefont {Grigorieva}, \citenamefont {Pokrovsky},\ and\
  \citenamefont {Atuchin}}]{Kalabin03p140}%
  \BibitemOpen
  \bibfield  {author} {\bibinfo {author} {\bibfnamefont {I.}~\bibnamefont
  {Kalabin}}, \bibinfo {author} {\bibfnamefont {T.~I.}\ \bibnamefont
  {Grigorieva}}, \bibinfo {author} {\bibfnamefont {L.~D.}\ \bibnamefont
  {Pokrovsky}}, \ and\ \bibinfo {author} {\bibfnamefont {V.~V.}\ \bibnamefont
  {Atuchin}},\ }in\ \href@noop {} {\emph {\bibinfo {booktitle} {Integrated
  Optical Devices: Fabrication and Testing}}},\ Vol.\ \bibinfo {volume} {4944}\
  (\bibinfo {organization} {International Society for Optics and Photonics},\
  \bibinfo {year} {2003})\ pp.\ \bibinfo {pages} {140--145}\BibitemShut
  {NoStop}%
\bibitem [{\citenamefont {Weller}\ and\ \citenamefont
  {Dickens}(1985)}]{Weller85p139}%
  \BibitemOpen
  \bibfield  {author} {\bibinfo {author} {\bibfnamefont {M.}~\bibnamefont
  {Weller}}\ and\ \bibinfo {author} {\bibfnamefont {P.}~\bibnamefont
  {Dickens}},\ }\href@noop {} {\bibfield  {journal} {\bibinfo  {journal}
  {Journal of Solid State Chemistry}\ }\textbf {\bibinfo {volume} {60}},\
  \bibinfo {pages} {139} (\bibinfo {year} {1985})}\BibitemShut {NoStop}%
\bibitem [{\citenamefont {Giannozzi}\ \emph {et~al.}(2009)\citenamefont
  {Giannozzi}, \citenamefont {Baroni}, \citenamefont {Bonini}, \citenamefont
  {Calandra}, \citenamefont {Car}, \citenamefont {Cavazzoni}, \citenamefont
  {Ceresoli}, \citenamefont {Chiarotti}, \citenamefont {Cococcioni},
  \citenamefont {Dabo}, \citenamefont {Corso}, \citenamefont {de~Gironcoli},
  \citenamefont {Fabris}, \citenamefont {Fratesi}, \citenamefont {Gebauer},
  \citenamefont {Gerstmann}, \citenamefont {Gougoussis}, \citenamefont
  {Kokalj}, \citenamefont {Lazzeri}, \citenamefont {Martin-Samos},
  \citenamefont {Marzari}, \citenamefont {Mauri}, \citenamefont {Mazzarello},
  \citenamefont {Paolini}, \citenamefont {Pasquarello}, \citenamefont
  {Paulatto}, \citenamefont {Sbraccia}, \citenamefont {Scandolo}, \citenamefont
  {Sclauzero}, \citenamefont {Seitsonen}, \citenamefont {Smogunov},
  \citenamefont {Umari},\ and\ \citenamefont
  {Wentzcovitch}}]{Giannozzi09p395502}%
  \BibitemOpen
  \bibfield  {author} {\bibinfo {author} {\bibfnamefont {P.}~\bibnamefont
  {Giannozzi}}, \bibinfo {author} {\bibfnamefont {S.}~\bibnamefont {Baroni}},
  \bibinfo {author} {\bibfnamefont {N.}~\bibnamefont {Bonini}}, \bibinfo
  {author} {\bibfnamefont {M.}~\bibnamefont {Calandra}}, \bibinfo {author}
  {\bibfnamefont {R.}~\bibnamefont {Car}}, \bibinfo {author} {\bibfnamefont
  {C.}~\bibnamefont {Cavazzoni}}, \bibinfo {author} {\bibfnamefont
  {D.}~\bibnamefont {Ceresoli}}, \bibinfo {author} {\bibfnamefont {G.~L.}\
  \bibnamefont {Chiarotti}}, \bibinfo {author} {\bibfnamefont {M.}~\bibnamefont
  {Cococcioni}}, \bibinfo {author} {\bibfnamefont {I.}~\bibnamefont {Dabo}},
  \bibinfo {author} {\bibfnamefont {A.~D.}\ \bibnamefont {Corso}}, \bibinfo
  {author} {\bibfnamefont {S.}~\bibnamefont {de~Gironcoli}}, \bibinfo {author}
  {\bibfnamefont {S.}~\bibnamefont {Fabris}}, \bibinfo {author} {\bibfnamefont
  {G.}~\bibnamefont {Fratesi}}, \bibinfo {author} {\bibfnamefont
  {R.}~\bibnamefont {Gebauer}}, \bibinfo {author} {\bibfnamefont
  {U.}~\bibnamefont {Gerstmann}}, \bibinfo {author} {\bibfnamefont
  {C.}~\bibnamefont {Gougoussis}}, \bibinfo {author} {\bibfnamefont
  {A.}~\bibnamefont {Kokalj}}, \bibinfo {author} {\bibfnamefont
  {M.}~\bibnamefont {Lazzeri}}, \bibinfo {author} {\bibfnamefont
  {L.}~\bibnamefont {Martin-Samos}}, \bibinfo {author} {\bibfnamefont
  {N.}~\bibnamefont {Marzari}}, \bibinfo {author} {\bibfnamefont
  {F.}~\bibnamefont {Mauri}}, \bibinfo {author} {\bibfnamefont
  {R.}~\bibnamefont {Mazzarello}}, \bibinfo {author} {\bibfnamefont
  {S.}~\bibnamefont {Paolini}}, \bibinfo {author} {\bibfnamefont
  {A.}~\bibnamefont {Pasquarello}}, \bibinfo {author} {\bibfnamefont
  {L.}~\bibnamefont {Paulatto}}, \bibinfo {author} {\bibfnamefont
  {C.}~\bibnamefont {Sbraccia}}, \bibinfo {author} {\bibfnamefont
  {S.}~\bibnamefont {Scandolo}}, \bibinfo {author} {\bibfnamefont
  {G.}~\bibnamefont {Sclauzero}}, \bibinfo {author} {\bibfnamefont {A.~P.}\
  \bibnamefont {Seitsonen}}, \bibinfo {author} {\bibfnamefont {A.}~\bibnamefont
  {Smogunov}}, \bibinfo {author} {\bibfnamefont {P.}~\bibnamefont {Umari}}, \
  and\ \bibinfo {author} {\bibfnamefont {R.~M.}\ \bibnamefont {Wentzcovitch}},\
  }\href@noop {} {\bibfield  {journal} {\bibinfo  {journal} {J. Phys.: Condens.
  Matter}\ }\textbf {\bibinfo {volume} {21}},\ \bibinfo {pages} {395502 (1}
  (\bibinfo {year} {2009})}\BibitemShut {NoStop}%
\bibitem [{\citenamefont {Rappe}\ \emph {et~al.}(1990)\citenamefont {Rappe},
  \citenamefont {Rabe}, \citenamefont {Kaxiras},\ and\ \citenamefont
  {Joannopoulos}}]{Rappe90p1227}%
  \BibitemOpen
  \bibfield  {author} {\bibinfo {author} {\bibfnamefont {A.~M.}\ \bibnamefont
  {Rappe}}, \bibinfo {author} {\bibfnamefont {K.~M.}\ \bibnamefont {Rabe}},
  \bibinfo {author} {\bibfnamefont {E.}~\bibnamefont {Kaxiras}}, \ and\
  \bibinfo {author} {\bibfnamefont {J.~D.}\ \bibnamefont {Joannopoulos}},\
  }\href@noop {} {\bibfield  {journal} {\bibinfo  {journal} {Phys. Rev. B Rapid
  Comm.}\ }\textbf {\bibinfo {volume} {41}},\ \bibinfo {pages} {1227} (\bibinfo
  {year} {1990})}\BibitemShut {NoStop}%
\bibitem [{\citenamefont {Perdew}\ \emph {et~al.}(1996)\citenamefont {Perdew},
  \citenamefont {Burke},\ and\ \citenamefont {Ernzerhof}}]{Perdew96p3865}%
  \BibitemOpen
  \bibfield  {author} {\bibinfo {author} {\bibfnamefont {J.~P.}\ \bibnamefont
  {Perdew}}, \bibinfo {author} {\bibfnamefont {K.}~\bibnamefont {Burke}}, \
  and\ \bibinfo {author} {\bibfnamefont {M.}~\bibnamefont {Ernzerhof}},\
  }\href@noop {} {\bibfield  {journal} {\bibinfo  {journal} {Phys. Rev. Lett.}\
  }\textbf {\bibinfo {volume} {77}},\ \bibinfo {pages} {3865 (1} (\bibinfo
  {year} {1996})}\BibitemShut {NoStop}%
\bibitem [{\citenamefont {Bravais}(1866)}]{Bravais86}%
  \BibitemOpen
  \bibfield  {author} {\bibinfo {author} {\bibfnamefont {A.}~\bibnamefont
  {Bravais}},\ }\href@noop {} {\bibfield  {journal} {\bibinfo  {journal}
  {Etudes Crisfallographiques}\ } (\bibinfo {year} {1866})}\BibitemShut
  {NoStop}%
\bibitem [{\citenamefont {Donnay}\ and\ \citenamefont
  {Harker}(1937)}]{Donnay37p446}%
  \BibitemOpen
  \bibfield  {author} {\bibinfo {author} {\bibfnamefont {J.}~\bibnamefont
  {Donnay}}\ and\ \bibinfo {author} {\bibfnamefont {D.}~\bibnamefont
  {Harker}},\ }\href@noop {} {\bibfield  {journal} {\bibinfo  {journal} {Am.
  Mineral}\ }\textbf {\bibinfo {volume} {22}},\ \bibinfo {pages} {446}
  (\bibinfo {year} {1937})}\BibitemShut {NoStop}%
\bibitem [{\citenamefont {Friedel}(1907)}]{Friedel07p326}%
  \BibitemOpen
  \bibfield  {author} {\bibinfo {author} {\bibfnamefont {G.}~\bibnamefont
  {Friedel}},\ }\href@noop {} {\bibfield  {journal} {\bibinfo  {journal} {Bull.
  Soc. Franc. Mineral}\ }\textbf {\bibinfo {volume} {07}},\ \bibinfo {pages}
  {326} (\bibinfo {year} {1907})}\BibitemShut {NoStop}%
\bibitem [{\citenamefont {Gonze}\ and\ \citenamefont
  {Lee}(1997)}]{Gonze97p10355}%
  \BibitemOpen
  \bibfield  {author} {\bibinfo {author} {\bibfnamefont {X.}~\bibnamefont
  {Gonze}}\ and\ \bibinfo {author} {\bibfnamefont {C.}~\bibnamefont {Lee}},\
  }\href@noop {} {\bibfield  {journal} {\bibinfo  {journal} {Phys. Rev. B}\
  }\textbf {\bibinfo {volume} {55}},\ \bibinfo {pages} {10355} (\bibinfo {year}
  {1997})}\BibitemShut {NoStop}%
\bibitem [{\citenamefont {Veithen}\ and\ \citenamefont
  {Ghosez}(2002)}]{Veithen02p214302}%
  \BibitemOpen
  \bibfield  {author} {\bibinfo {author} {\bibfnamefont {M.}~\bibnamefont
  {Veithen}}\ and\ \bibinfo {author} {\bibfnamefont {P.}~\bibnamefont
  {Ghosez}},\ }\href@noop {} {\bibfield  {journal} {\bibinfo  {journal} {Phys.
  Rev. B}\ }\textbf {\bibinfo {volume} {65}},\ \bibinfo {pages} {214302}
  (\bibinfo {year} {2002})}\BibitemShut {NoStop}%
\bibitem [{\citenamefont {Rohlfing}\ and\ \citenamefont
  {Louie}(2000)}]{Rohlfing00p4927}%
  \BibitemOpen
  \bibfield  {author} {\bibinfo {author} {\bibfnamefont {M.}~\bibnamefont
  {Rohlfing}}\ and\ \bibinfo {author} {\bibfnamefont {S.~G.}\ \bibnamefont
  {Louie}},\ }\href@noop {} {\bibfield  {journal} {\bibinfo  {journal} {Phys.
  Rev. B.}\ }\textbf {\bibinfo {volume} {62}},\ \bibinfo {pages} {4927}
  (\bibinfo {year} {2000})}\BibitemShut {NoStop}%
\bibitem [{\citenamefont {Zelmon}\ \emph {et~al.}(1997)\citenamefont {Zelmon},
  \citenamefont {Small},\ and\ \citenamefont {Jundt}}]{Zelmon97p3319}%
  \BibitemOpen
  \bibfield  {author} {\bibinfo {author} {\bibfnamefont {D.}~\bibnamefont
  {Zelmon}}, \bibinfo {author} {\bibfnamefont {D.}~\bibnamefont {Small}}, \
  and\ \bibinfo {author} {\bibfnamefont {D.}~\bibnamefont {Jundt}},\
  }\href@noop {} {\bibfield  {journal} {\bibinfo  {journal} {JOSA B}\ }\textbf
  {\bibinfo {volume} {14}},\ \bibinfo {pages} {3319} (\bibinfo {year}
  {1997})}\BibitemShut {NoStop}%
\end{thebibliography}%

\end{document}